%% file: 9399.tex
\documentclass{aa}  
\pdfoutput=1
\usepackage{graphicx}
\usepackage{txfonts}
\usepackage{natbib}
\usepackage{longtable}
\usepackage{ifpdf}
\bibpunct{(}{)}{;}{a}{}{,} % to follow A&A style
\begin{document}

  \title{The Araucaria Project: VLT-spectroscopy of blue massive stars in NGC 55\thanks{Based on observations obtained at 
  the ESO VLT for Large Programme 171.D-0004.}}
%\titlerunning

   \author{N. Castro \inst{1,2}
          \and 
 	    A. Herrero \inst{1,2}
           \and 
	    M. Garcia \inst{1} 
          \and 
	    C. Trundle \inst{3}
          \and 
	    F. Bresolin \inst{4}
          \and 
	    W. Gieren \inst{5} \\
          \and 
            G. Pietrzy{\'n}ski \inst{5,6}			
          \and 
	    R.~-P. Kudritzki \inst{4}
	  \and
	    R. Demarco \inst{7}    
	  }

   \offprints{N. Castro (\email{norberto@iac.es})}

  \institute{Instituto de Astrof\'{i}sica de Canarias, C/ V\'{i}a L\'{a}ctea s/n, E-38200 La Laguna, Tenerife, Spain.
         \and
	 	Departamento de Astrof\'{i}sica, Universidad de La Laguna, Avda. Astrof\'{i}sico
		Francisco S\'anchez s/n, E-38071 La Laguna, Tenerife, Spain.
	 \and
	 	Astronomy Research Centre, Department of Physics \& Astronomy, School of Mathematics \& Physics,
		The Queen's University of Belfast, Belfast, Northern Ireland, UK.
	\and
		Institute for Astronomy, 2680 Woodlawn Drive, Honolulu, HI 96822, USA.
	\and
		Departamento de F\'{i}sica, Astronomy Group, Universidad de Concepci\'on, Casilla 160-C, Concepci\'on, Chile.
	\and
		Warsaw University Observatory, Al. Ujazdowskie 4,00-478, Warsaw, Poland. 
	\and
		Department of Physics and Astronomy, The Johns Hopkins University, Baltimore, MD 21218, USA.
     }

   \date{Received January 15, 2008; accepted March 31, 2008}

% \abstract{}{}{}{}{} 
% 5 {} token are mandatory
 
  \abstract
{}
{This is the first paper of a series devoted to studying the population of blue massive stars in NGC 55, a galaxy of the 
Sculptor group at a distance of about 2 Mpc.}
{We have obtained optical ($3300-6210 \AA$), low-resolution spectra of approximately 200 blue massive stars with VLT-FORS2, 
which we have classified with the aid of Milky Way and Magellanic Cloud standard stars.}
{We present the first census of massive blue stars in NGC 55. A study of stellar radial velocities shows agreement with 
existing \ion{H}{I} rotational velocity curve work and reveals the presence of one object with peculiar velocity. A 
qualitative study of the stellar metallicity suggests that its global distribution over NGC 55 is close to that of the 
LMC, as derived from previous studies.}
{We present a catalogue with 164 classifications of blue massive stars in NGC 55. This catalogue is a first and necessary 
step for the subsequent quantitative study of blue massive stars in NGC 55 with state-of-the-art model atmospheres.}

  
	\authorrunning{N. Castro et al.}
	\titlerunning{Massive stars in NGC 55}
   \maketitle

%%%%%%%%%%%%%%%%%%%%%%%%%%%%%%%%%%%%%%%%%%%%%%%%%%%%%%%%%%%%%%%%%%%%%%%%%%%%%%%%%%%%%%%%
%%%%%%%%%%%%%%%%%%%%%%%%%%%%%%%%%%%%%%%%%%%%%%%%%%%%%%%%%%%%%%%%%%%%%%%%%%%%%%%%%%%%%%%%

\section{INTRODUCTION}

   \begin{figure*}
   \resizebox{\hsize}{!}{\includegraphics[angle=90]{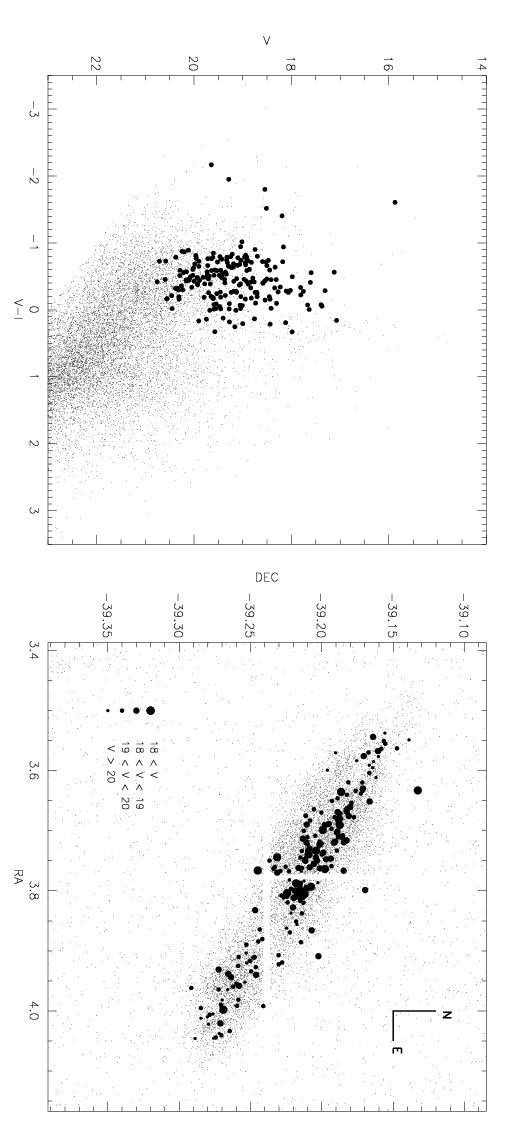}}	
      \caption{Left: Colour-magnitude diagram of NGC 55. The targets selected for the spectroscopic 
      study presented in this paper are marked with circles. Right: Location of the target stars in NGC 55. 
      Note that they are spread across the galaxy providing information from different points of NGC 55.}
         \label{Fig:CM}
   \end{figure*}

The exploration of massive stars in nearby galaxies is experiencing a renewed interest since it became possible to 
study them spectroscopically outside the Milky Way, even beyond the Local Group, with the new generation of 8-10m class 
telescopes. There are many reasons for our interest in massive stars: they are the main source of ionisation and chemical 
enrichment of the Universe and their influence on the chemical and dynamical evolution of galaxies is very strong due to 
their stellar winds and supernova explosions. Additionally, blue massive stars can be used to measure distance by 
means of the relationship between their luminosity and their wind momentum \citep{1995svlt.conf..246K} or their 
flux-weighted gravity $g/T_{eff}^4$ \citep{2003ApJ...582L..83K,2008ApJ...999...999}. Once calibrated, these techniques 
can provide distance estimates to the Virgo and Fornax clusters as an alternative distance diagnostic to Cepheids, and 
with comparable accuracy (see \citealt{1999A&A...350..970K} and references above). However, our knowledge of massive stars 
is incomplete and the physics involved in their formation and evolution is not totally understood 
\citep{2000ARA&A..38..613K,2003ARA&A..41...15M}.

Modelling the atmospheres of these stars is a complex task involving non-local thermodynamical equilibrium processes, 
and strong stellar winds in a spherically extended atmosphere. Each atmospheric model is computed with several 
free parameters, all of which have different effects on the emerging spectrum. The stellar metal content is  
crucial, since the stellar wind is driven by the moment transfer between photons escaping from the photosphere 
and metallic ions \citep{2000ARA&A..38..613K}. Beyond the Milky Way we can study the impact of metallicity 
on stellar properties, evolution and feedback. Consequently, there is growing interest in studies of massive 
stars in galaxies with different metal content. In our Galaxy there are numerous studies of clusters and 
associations \citep{1999A&A...348..542H,2000A&A...363..537R,2001A&A...367...86S,2002A&A...396..949H,
2004ApJ...617.1115D,2006A&A...457..265D}, but the Magellanic Clouds are the most targeted due to their proximity 
and their different metallicity from the Milky Way \citep{2003A&A...398..455L,2003A&A...400...21R,2004A&A...417..217T,
2005ApJ...633..899K,2005A&A...438..265L,2005A&A...434..677T,2007A&A...466..277H,2007A&A...465.1003M,2007A&A...471..625T}. 
Additional work on other galaxies of the Local Group include: M 31 \citep{2000ApJ...541..610V,2002A&A...395..519T}, 
M 33 \citep{1998Ap&SS.263..171M,2000ApJ...545..813M,2005ApJ...635..311U}, NGC 6822 
\citep{1999A&A...352L..40M,2001ApJ...547..765V}, NGC 3109 \citep{2007ApJ...659.1198E}, WLM 
\citep{2003AJ....126.1326V,2006ApJ...648.1007B} or IC 1613 \citep{2007ApJ...671.2028B}. Detailed quantitative analyses 
of extragalactic massive stars reach as far as NGC 300 in the Sculptor group \citep{2002ApJ...567..277B,2005ApJ...622..862U} 
where iron-group elements have been studied for the first time outside the Local Group by \cite{2008ApJ...999...999}. 
Beyond that, \cite{2001ApJ...548L.159B} managed to determine global metallicities for A-type supergiants in NGC 3621 
at a distance of 6.7 Mpc.

We now extend the study of massive extragalactic stars to NGC 55. This galaxy is located in the Sculptor group at 1.94 Mpc 
\citep{2006AJ....132.2556P,2007arXiv0709.2421G}, close enough to allow quantitative spectroscopic analyses of bright 
blue stars. Its large inclination angle ($\sim$80$\degr$, \citealt{1986A&A...166...97H}) makes its morphological 
classification difficult. Some authors argue that NGC~55 is a Magellanic irregular (see for instance 
\citealt{2005ApJ...622..279D}), however we adopt \cite{1961ApJ...133..405D}'s SB(s)m classification for this work. 
It is one of the largest galaxies of the Sculptor group, together with NGC 300 and NGC 247, and its chemical composition 
is similar to that of the Large Magellanic Cloud (LMC) ([\ion{Fe}/\ion{H}]$\sim$$-0.3$,  \citealt{2005ApJ...622..279D}). 
Photometric studies hint an important population of blue stars \citep{2005Msngr.121...23G} mainly in the central 
region, where intense stellar activity is revealed by bubbles and filaments produced by strong stellar winds and supernova 
explosions \citep{2003A&A...412...69T}.

In this paper the first study and census of massive blue stars in NGC 55 is presented. In section 2 we describe briefly 
the target selection process, the observations and the data reduction. In section 3 we discuss  the spectral 
classification criteria for our catalogue of blue stars in NGC 55. We provide rough estimates for the distribution 
of stellar radial velocities and metallicities in sections 4 and 5. Finally, a summary and our conclusions are presented 
in section 6.

   \begin{figure}
   \resizebox{\hsize}{!}{\includegraphics[angle=0]{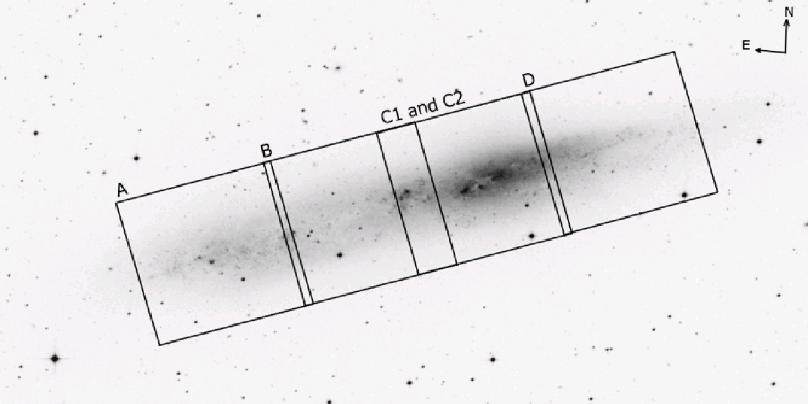}}	
     \caption{Observed fields of NGC 55. 
      Each square represents approximately the FORS2 field of view ($ 6'.8\times 6'.8$) in MXU mode. The image was taken 
      from the {\it{DSS}} archive ($http://archive.stsci.edu/cgi-bin/dss\_form$).}
         \label{Fig1}
   \end{figure}

%%%%%%%%%%%%%%%%%%%%%%%%%%%%%%%%%%%%%%%%%%%%%%%%%%%%%%%%%%%%%%%%%%%%%%%%%%%%%%%%%%%%%%%%
%%%%%%%%%%%%%%%%%%%%%%%%%%%%%%%%%%%%%%%%%%%%%%%%%%%%%%%%%%%%%%%%%%%%%%%%%%%%%%%%%%%%%%%%

\section{OBSERVATIONS  AND DATA REDUCTION}

\begin{table*}
\caption{VLT-FORS2 observations of NGC 55.}
\label{table_1}
\centering
\begin{tabular}{c c c c c}
\hline
\hline
Field & RA (J2000) & DEC (J2000)      & Total T$_{exp}$ (s)   &  Date \\
\hline
A      &00:15:55.0  &-39:16:13.0      &10800             &2004-11-07	      \\
B      &00:15:22.9  &-39:13:54.6      &13500  		 &2004-11-10	      \\
C1     &00:14:57.6  &-39:12:28.5      &10800  		 &2004-11-06	      \\
C2     &00:14:57.6  &-39:12:28.5      &10800  		 &2004-11-06	      \\
D      &00:14:25.5  &-39:10:21.4      &16200   		 &2004-11-07/2004-11-10  \\
\hline
\end{tabular}
\end{table*}

Targets were selected from existing V- and I-band photometry obtained as part of the ongoing ARAUCARIA Cepheid search 
project\footnote{http://ezzelino.ifa.hawaii.edu/$\sim$bresolin/Araucaria/index.html} whose results have been reported in 
\cite{2006AJ....132.2556P}. These data were taken with the Warsaw 1.3 m telescope at Las Campanas Observatory. The 
telescope is equipped with a mosaic CCD $8k\times 8k$  with a field of view of  $\sim35'\times 35'$ and a scale of 
about $\sim0.25$ ''/pix. The galaxy's distance modulus is $m-M=26.43$ \citep{2007arXiv0709.2421G}. Accordingly, O and 
early-B supergiants with $M_v\sim-6.5$  have an apparent magnitude $V\lesssim 20$, sufficient for low-resolution 
spectroscopy.  Magnitude and colour $(V-I)\lesssim 0.0$ were the main criteria for candidate selection. NGC 55 has a 
high galactic latitude ($b=-75\degr$) and the foreground reddening is low ($E(B-V)=0.013$, \citealt{1998ApJ...500..525S}). 
Nonetheless, the internal reddening of the galaxy may not be negligible since NGC 55 is almost "edge-on". In fact, the 
internal reddening affecting the Cepheids in NGC 55 has been estimated from a multiwavelength analysis in 
\cite{2007arXiv0709.2421G} to be $E(B-V)=0.13$. Figure \ref{Fig:CM} shows the colour-magnitude diagram from which targets 
were selected, along with a plot identifying their positions in the galaxy.

The final list of candidates was built by careful  examination of the images to reject objects with nearby companions 
(real or projected) and to avoid overlap with \ion{H}{II} regions, a hard task given the inclination of the galaxy. We 
cannot eliminate the possibility that the targets are unresolved binaries, since at the distance of NGC 55 one arc-second 
corresponds to $\sim 9.2$ pc. In fact, the occurrence of binarity in blue massive stars is quite high. According to 
\cite{1998AJ....115..821M} in their study 72.2\% of O-type stars in clusters are in binary systems, while the rate 
is 19.4\% for field stars. In addition, the low spectral resolution of our data makes the detection of binaries from 
their spectra a difficult task.

We obtained spectra for $\sim$200 objects with FORS2 (FOcal Reducer and low dispersion Spectrograph, 
\cite{1998Msngr..94....1A}) at the Very Large Telescope (VLT-UT2), in the Mask eXchange Unit (MXU) mode. The observations, 
performed in November 2004, are summarised in Table \ref{table_1}. The detector is a mosaic of two $2k \times 4k$ MIT CCDs 
with a gap of $4"$. The total field of view is $6'.8\times 6'.8$. Four pointings were necessary to cover the whole galaxy, 
as shown in Fig. \ref{Fig1}. One mask per field was built with FIMS (FORS Instrumental Mask Simulator), except for the 
central pointing where the population of blue massive stars is expected to be dense and two masks were needed. The spectra 
were taken with the 600B grism which provided the highest resolving power for this study ($\lambda/\Delta\lambda=780$) 
in the wavelength range of interest ($3100-6210$\AA). Each slit was carved with a width of 1 arc-second. The position 
of the target stars studied in this paper marked on the FORS2 V-band preimaging is shown in Figs. \ref{FindA} 
to \ref{FindD}.

The spectra were reduced with the pipeline developed by \cite{2005A&A...432..381D} optimised for FORS2 data and upgraded 
for this work with the implementation of two new modules: \cite{2001PASP..113.1420V}'s algorithm for cosmic ray rejection, 
based on the laplacian edge detection; and a routine to normalise flat field images prior to performing the flat-field 
correction. The pipeline extracts all slits and works with them separately, taking into account the deformation of the 
slits on the CCD. The code then follows the standard steps for long slit spectrum reduction, using IRAF\footnote{IRAF is 
distributed by the National Optical Astronomy Observatory, which is operated by the Association of Universities for 
Research in Astronomy, Inc., under cooperative agreement with  the National Science Foundation.} 
tasks: bias, flat-field correction and background subtraction. The latter was the most challenging one, since in some 
instances we were unable to completely correct for nebular contamination. This was particularly problematic in the case 
of some of the Balmer lines which were filled with nebular emission (see below). The pipeline produces one-dimensional 
spectra with an optimised wavelength calibration. Finally, the individual exposures were combined (weighted by their 
count number) and rectified to the local continuum (normalised).

Sky subtraction was a laborious task due to nebular emission, which is particularly strong in the central regions of the 
galaxy. This contamination is more severe in young stars, which are still embedded in the dense regions where 
they were born. Interstellar absorption lines (\ion{Ca}{II}$\lambda3933-3968$) can also contaminate the spectrum 
\citep{1904ApJ....19..268H}. Moreover, given the high inclination of the galaxy, photons from a large number of 
\ion{H}{II} regions (with different radial velocities) are integrated along the line of sight. One last problem in the 
sky subtraction is the low spectral resolution of the data. In spite of all our efforts, the spectra of several stars are  
strongly contaminated by the background. Nevertheless, in the vast majority of our 
targets the nebular lines do not compromise the spectral classification.

The reduced spectra have a signal-to-noise ratio (S/N) of $\sim75$ on average, sufficient for spectral classification. 
The spectra of the brightest objects with S/N $\geq100$ can be analysed quantitatively, which will be done in a future 
paper. A few objects are underexposed with  S/N smaller than 50.

%%%%%%%%%%%%%%%%%%%%%%%%%%%%%%%%%%%%%%%%%%%%%%%%%%%%%%%%%%%%%%%%%%%%%%%%%%%%%%%%%%%%%%%%
%%%%%%%%%%%%%%%%%%%%%%%%%%%%%%%%%%%%%%%%%%%%%%%%%%%%%%%%%%%%%%%%%%%%%%%%%%%%%%%%%%%%%%%%

\section{SPECTRAL CLASSIFICATION}

\begin{table}
\caption{Number of NGC 55 blue massive stars identified in this work, separated by field and spectral type.}
\label{table_num}
\centering
\begin{tabular}{llllll}
\hline
\hline
  Field  & O-Type     & B-Type        & A-Type & WR-LBVc &  Total \\
\hline
A	& 5  & 20 & 13 & 1  &  39  \\
B	& 4  & 7  & 7  & 3  &  21  \\
C	& 4  & 34 & 30 & 3  &  71  \\
D	& 1  & 14 & 18 & 0  &  33  \\
Total	& 14 & 75 & 68 & 7  & 164  \\
\hline
\end{tabular}
\end{table}

%%%%%%%%%%%%%%%%%%%%%%%%%%%%%%%%%%%%%%%%%%%%%%%%%%%%%%%%%%%%%%

\begin{table*}
\caption{A subset
 of blue massive stars showing a variety of spectral types (O-, B-, A-types plus WRs and LBV candidates, LBVc) 
observed in NGC 55.} 
\label{catalog1}
\centering
\begin{tabular}{ccccclccccl}
\hline\hline
\hline
%\endfirsthead
 \scriptsize{ID}  & \scriptsize{ RA(J2000)} &  \scriptsize{DEC(J2000)}  &  \scriptsize{V}  &  \scriptsize{V-I}  & \scriptsize{Spectral Type} & \scriptsize{Rg} & \scriptsize{$v_r$}  &  \scriptsize{S/N}   &  \scriptsize{Metallicity}  &  \scriptsize{Comments}  \\
 \scriptsize{(1)}  & \scriptsize{(2)} &  \scriptsize{(3)}  &  \scriptsize{(4)}  &  \scriptsize{(5)}  & \scriptsize{(6)} & \scriptsize{(7)} & \scriptsize{(8)} &  \scriptsize{(9)}  &  \scriptsize{(10)}  &  \scriptsize{(11)}  \\
\hline
 \scriptsize{   D\_36}  &  \scriptsize{  0:14:35.07}  &  \scriptsize{-39:11:20.04}  &  \scriptsize{19.189}  &  \scriptsize{-0.689}  &  \scriptsize{        B9I}  &  \scriptsize{ -278.26}  &  \scriptsize{  78}  &  \scriptsize{  70}  &  \scriptsize{     LMC}  &  \scriptsize{  	  }  \\
 \scriptsize{   C2\_8}  &  \scriptsize{  0:14:43.87}  &  \scriptsize{-39:12:29.51}  &  \scriptsize{19.294}  &  \scriptsize{-0.652}  &  \scriptsize{        A0I}  &  \scriptsize{ -147.56}  &  \scriptsize{  95}  &  \scriptsize{  69}  &  \scriptsize{     SMC}  &  \scriptsize{  	  }  \\
 \scriptsize{  C1\_18}  &  \scriptsize{  0:14:55.94}  &  \scriptsize{-39:11:21.84}  &  \scriptsize{19.316}  &  \scriptsize{-0.707}  &  \scriptsize{        WN8}  &  \scriptsize{  -37.95}  &  \scriptsize{ 124}  &  \scriptsize{  30}  &  \scriptsize{	 -}  &  \scriptsize{  	      neb.}  \\
 \scriptsize{  C1\_30}  &  \scriptsize{  0:14:59.91}  &  \scriptsize{-39:12:11.88}  &  \scriptsize{18.921}  &  \scriptsize{-0.425}  &  \scriptsize{       LBVc}  &  \scriptsize{   95.62}  &  \scriptsize{ 158}  &  \scriptsize{  28}  &  \scriptsize{	 -}  &  \scriptsize{              }  \\
 \scriptsize{  C1\_37}  &  \scriptsize{  0:15:02.50}  &  \scriptsize{-39:13:58.44}  &  \scriptsize{19.642}  &  \scriptsize{-0.588}  &  \scriptsize{       A7II}  &  \scriptsize{  153.43}  &  \scriptsize{ 146}  &  \scriptsize{  55}  &  \scriptsize{	 -}  &  \scriptsize{  		  }  \\
 \scriptsize{  C1\_51}  &  \scriptsize{  0:15:14.70}  &  \scriptsize{-39:12:54.36}  &  \scriptsize{18.306}  &  \scriptsize{-0.098}  &  \scriptsize{   Ofpe/WN9}  &  \scriptsize{  319.21}  &  \scriptsize{ 187}  &  \scriptsize{  57}  &  \scriptsize{	 -}  &  \scriptsize{  	      neb.}  \\
 \scriptsize{   B\_12}  &  \scriptsize{  0:15:16.68}  &  \scriptsize{-39:13:26.40}  &  \scriptsize{19.108}  &  \scriptsize{-0.666}  &  \scriptsize{      Of/WN}  &  \scriptsize{  351.60}  &  \scriptsize{ 176}  &  \scriptsize{  90}  &  \scriptsize{	 -}  &  \scriptsize{  	      neb.}  \\
 \scriptsize{   B\_13}  &  \scriptsize{  0:15:18.63}  &  \scriptsize{-39:13:12.72}  &  \scriptsize{18.543}  &  \scriptsize{-0.435}  &  \scriptsize{  LBVc/WN11}  &  \scriptsize{  379.26}  &  \scriptsize{ 274}  &  \scriptsize{ 103}  &  \scriptsize{	 -}  &  \scriptsize{              }  \\
 \scriptsize{   B\_34}  &  \scriptsize{  0:15:37.73}  &  \scriptsize{-39:13:49.08}  &  \scriptsize{19.533}  &  \scriptsize{-0.242}  &  \scriptsize{  LBVc/WN11}  &  \scriptsize{  667.26}  &  \scriptsize{ 216}  &  \scriptsize{  35}  &  \scriptsize{	 -}  &  \scriptsize{              }  \\
 \scriptsize{   B\_36}  &  \scriptsize{  0:15:40.62}  &  \scriptsize{-39:13:40.07}  &  \scriptsize{19.782}  &  \scriptsize{-0.389}  &  \scriptsize{        O8I}  &  \scriptsize{  709.64}  &  \scriptsize{ 171}  &  \scriptsize{  55}  &  \scriptsize{	 -}  &  \scriptsize{  		  }  \\
 \scriptsize{    A\_5}  &  \scriptsize{  0:15:40.83}  &  \scriptsize{-39:15:09.72}  &  \scriptsize{20.071}  &  \scriptsize{-0.495}  &  \scriptsize{        A3I}  &  \scriptsize{  722.43}  &  \scriptsize{ 260}  &  \scriptsize{  58}  &  \scriptsize{	 -}  &  \scriptsize{  		  }  \\
 \scriptsize{    A\_8}  &  \scriptsize{  0:15:44.18}  &  \scriptsize{-39:14:58.20}  &  \scriptsize{19.792}  &  \scriptsize{-0.188}  &  \scriptsize{      O9.7I}  &  \scriptsize{  770.70}  &  \scriptsize{ 304}  &  \scriptsize{  62}  &  \scriptsize{	 -}  &  \scriptsize{  		  }  \\
 \scriptsize{   A\_11}  &  \scriptsize{  0:15:45.17}  &  \scriptsize{-39:15:56.15}  &  \scriptsize{18.706}  &  \scriptsize{-0.245}  &  \scriptsize{        B5I}  &  \scriptsize{  793.59}  &  \scriptsize{ 209}  &  \scriptsize{ 130}  &  \scriptsize{     LMC}  &  \scriptsize{          }  \\
 \scriptsize{   A\_29}  &  \scriptsize{  0:15:57.97}  &  \scriptsize{-39:15:33.84}  &  \scriptsize{19.575}  &  \scriptsize{-0.595}  &  \scriptsize{      O5If+}  &  \scriptsize{  980.35}  &  \scriptsize{ 210}  &  \scriptsize{  62}  &  \scriptsize{	 -}  &  \scriptsize{  	     neb. }  \\
 \scriptsize{   A\_34}  &  \scriptsize{  0:16:01.17}  &  \scriptsize{-39:16:34.67}  &  \scriptsize{20.247}  &  \scriptsize{-0.875}  &  \scriptsize{        B0I}  &  \scriptsize{ 1036.66}  &  \scriptsize{ 215}  &  \scriptsize{  68}  &  \scriptsize{     LMC}  &  \scriptsize{          }  \\
 \scriptsize{   A\_42}  &  \scriptsize{  0:16:09.69}  &  \scriptsize{-39:16:13.44}  &  \scriptsize{19.453}  &  \scriptsize{-0.189}  &  \scriptsize{       LBVc}  &  \scriptsize{ 1160.17}  &  \scriptsize{ 175}  &  \scriptsize{  75}  &  \scriptsize{	 -}  &  \scriptsize{              }  \\
\hline
\hline
\end{tabular}
%~\\
\note{The complete catalogue is gathered in Table \ref{t_catalog}.}
%~\\
\end{table*}

%%%%%%%%%%%%%%%%%%%%%%%%%%%%%%%%%%%%%%%%%%%%%%%%%%%%%%%%%%%%%%
%%%%%%%%%%%%%%%%%%%%%%%%%%%%%%%%%%%%%%%%%%%%%%%%%%%%%%%%%%%%%%

   \begin{figure*}
   \resizebox{\hsize}{!}{\includegraphics[angle=0]{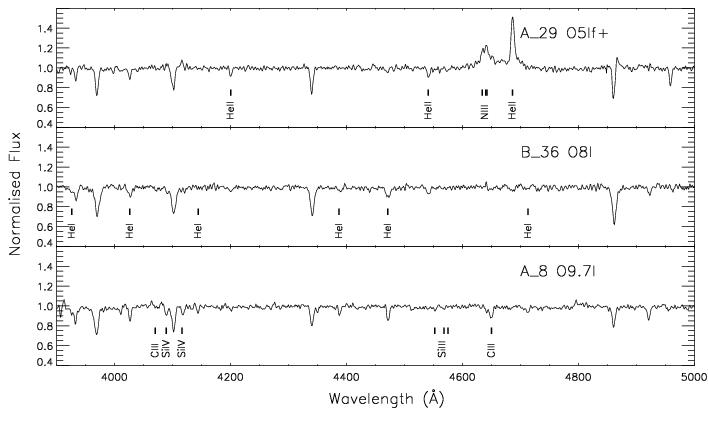}}	
     \caption{O-type supergiants in NGC 55. The most relevant transitions for spectral classification are shown.}
	\label{Fig:otipo}
  \end{figure*}

%%%%%%%%%%%%%%%%%%%%%%%%%%%%%%%%%%%%%%%%%%%%%%%%%%%%%%%%%%

  \begin{figure*}
   \resizebox{\hsize}{!}{\includegraphics[angle=0]{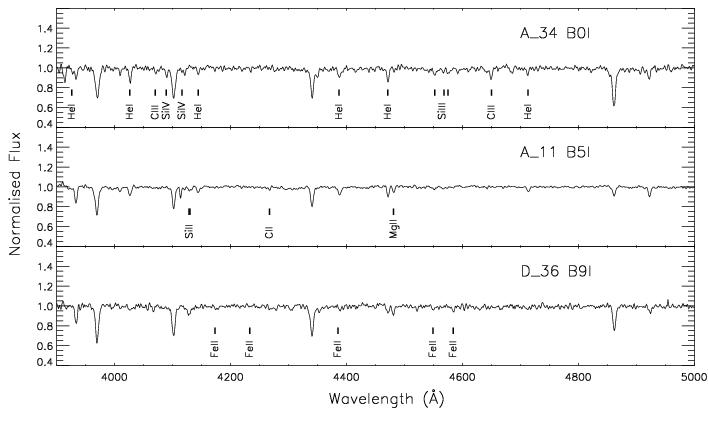}}	
      \caption{Optical spectra of B-type supergiants in NGC 55.}
         \label{Fig:btipo}
  \end{figure*}

%%%%%%%%%%%%%%%%%%%%%%%%%%%%%%%%%%%%%%%%%%%%%%%%%%%%%%

  \begin{figure*}
    \resizebox{\hsize}{!}{\includegraphics[angle=0]{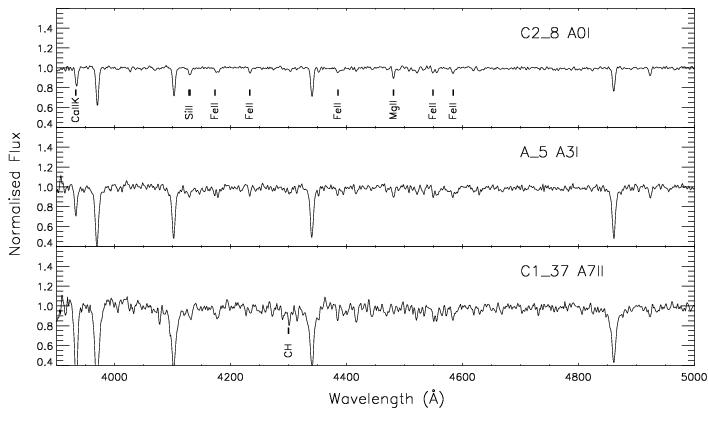}}	
   
     \caption{Optical spectra of A-supergiants in NGC 55.}
	\label{Fig:atipo}
  \end{figure*}

%%%%%%%%%%%%%%%%%%%%%%%%%%%%%%%%%%%%%%%%%%%%%%%%%%%%%%

  \begin{figure*}
   \resizebox{\hsize}{!}{\includegraphics[angle=0]{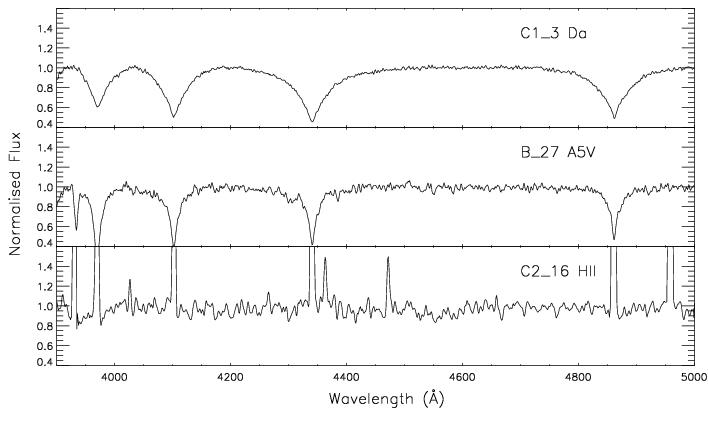}}	
  
     \caption{Three exceptional cases: a white dwarf (top), an A5V detected in our sample probably 
     belonging to the Milky Way halo (middle) and an example of spectra dominated by \ion{H}{II} emission (bottom).}
	\label{enana}
  \end{figure*}

%%%%%%%%%%%%%%%%%%%%%%%%%%%%%%%%%%%%%%%%%%%%%%%%%%%%%%

   \begin{figure*}
   \resizebox{\hsize}{!}{\includegraphics[angle=0]{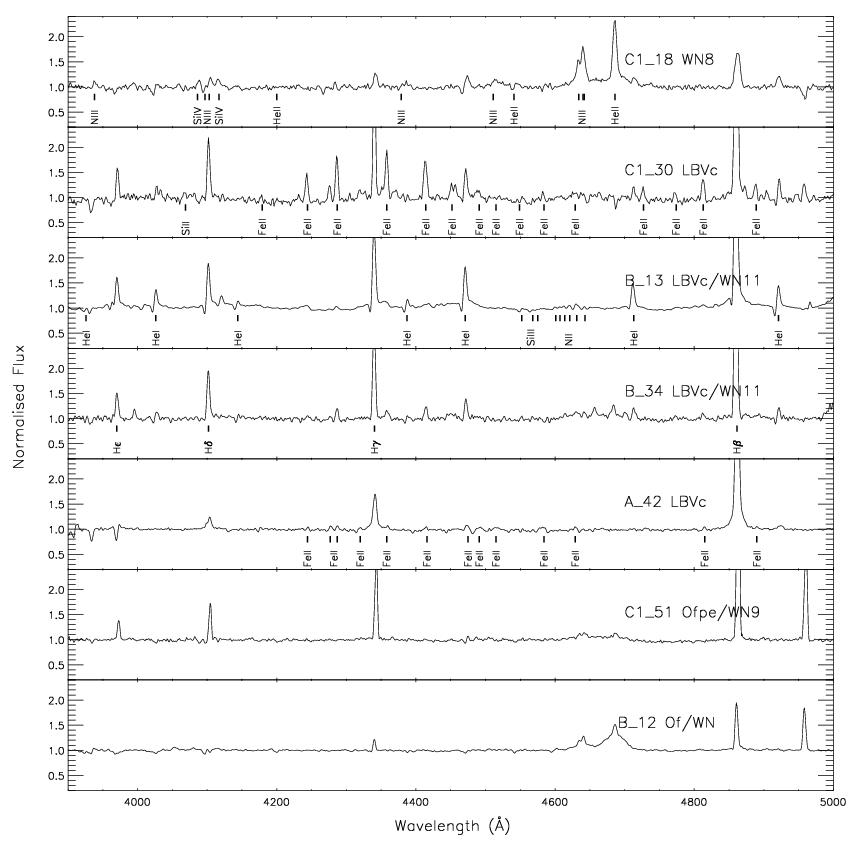}}	
      \caption{"Exotic" stars found in our survey of NGC 55. See Sect. \ref{sec:wr} for a discussion
 of the morphology of each object.}
         \label{Fig:wrtipo}
   \end{figure*}

The morphological analysis of stellar spectra is a crucial first step in studying the population of massive stars in a 
galaxy and provides clues about the stellar parameters of the stars. The main advantage of a classification system is 
that it preserves its independence from theoretical models or other external information \citep{1979RA......9..337W}.

The system developed by \cite{1943QB881.M6.......}, together with subsequent refinements by \cite{1990PASP..102..379W} 
(hereafter WF) and \cite{1992A&AS...94..569L}, is the basis for the classification of Galactic stars. However, 
metallicity alters the strength of the metallic diagnostic lines and therefore the stellar classification criteria 
are dependent on the star's metallicity. In newly explored galaxies where the metallicity is not accurately known, we 
have to be cautious and use the appropriate standards, since the chemical composition of the studied stars may differ 
from all existing templates. In light of this, we also used as standards those of \cite{1991PASP..103.1123F} from the LMC 
and \cite{1997A&A...317..871L} and \cite{2003MNRAS.345.1223E} from the Small Magellanic Cloud (SMC).

NGC 55 seems to have a metallicity similar to the LMC \citep{2003A&A...412...69T,2005ApJ...622..279D}. Following the idea 
of \cite{1991PASP..103.1123F}, we have performed the spectral classification of our sample by  visual comparison with 
templates from LMC stars. Even though the metallicities of both galaxies are similar, we have been careful with criteria 
involving both helium or hydrogen lines and metallic transitions. Criteria based only on metallic transitions, for 
instance \ion{Si}{III}/\ion{Si}{IV}, were preferred when available. The luminosity class was determined following  WF for 
O-type stars and using the width of Balmer lines for B- and A-stars \citep{1987A&AS...69..421A}. We also took into account 
that luminosity and spectral type are not independent. The resolution and quality of the spectra prevented us from doing 
a luminosity classification finer than I, II, III and V. Besides, the lack of H$\alpha$ spectra and the nebular 
contamination further impairs this distinction. Additional diagnostic metallic lines for luminosity class of B- and 
A-types exist (see for instance WF), but the variations are smaller given the low metallicity of NGC 55 and cannot be 
detected at the quality and resolution of the spectra.

We have provided detailed classifications for 204 spectra, out of which 164 stars seem to belong to NGC 55 and have 
spectral types earlier than F0. The rest of the objects include F- and G-supergiants (23 stars), \ion{H}{II} regions 
(4 spectra) and a sample of B-, A-, F-, and G-dwarfs (13 stars) plus a white dwarf. The number of different types of 
blue massive stars found in NGC 55 is provided in Table \ref{table_num}. The complete catalogue is available in the 
 appendices. Table \ref{t_catalog} lists the set of early-type stars (earlier than F0) found in NGC 55 (in Table 
\ref{catalog1} we show a fragment of the total database as an illustration, including some objects presented in following 
figures). The catalogue columns are: (1) star identification, consisting of field number followed by slit number; (2) 
right ascension (hh:mm:ss); (3) declination (dd:mm:ss); (4) V-magnitude; (5) V-I colour; (6) spectral type; (7) projected 
galactocentric distance (arc-seconds); (8) radial velocity (km s$^{-1}$); (9) signal-to-noise ratio of the spectrum; (10) 
metallicity estimate; (11) comments. The corresponding digital spectral atlas is displayed in Figs. 
\ref{ape0} -- \ref{ape20}. F- and G-stars (classified using \citealt{1990clst.book.....J} and templates from 
\citealt{2003A&A...402..433L}) are compiled in Table \ref{t_catalogF}, for completeness. Finally, Table \ref{t_catalogMW} 
lists dwarf stars together with a Galactic white dwarf found in the sample.

The accuracy of the  classification depends on several factors. It is essential to have spectra of sufficiently high S/N. 
Faint stars have been assigned only a very rough classification. A good subtraction of cosmic rays and a complete sky 
subtraction are key points. The latter was not always possible, leaving the Balmer lines (main luminosity class indicators) 
contaminated by nebular emission at their cores. However, the impact on the luminosity classification was negligible in 
most cases since we could use the Balmer line wings and the relevant metallic lines were not affected. We have marked the 
spectra with remaining nebular contamination in the Balmer lines or in the [\ion{O}{III}] lines (about 46\% of the sample) 
with comment {\it{'neb.'}} in Tables \ref{catalog1}, \ref{t_catalog}, \ref{t_catalogF} and \ref{t_catalogMW}.

We have found a total of 14 O-type stars in our sample. Fig. \ref{Fig:otipo} shows a subset of these objects. Most of the 
observed stars (75 of them) are B-type, some examples are presented in  Fig. \ref{Fig:btipo}. We have classified 68 
stars as A-type and some representative examples can be seen in Fig. \ref{Fig:atipo}. In  these figures we have marked the 
most important lines used in the spectral classification, following the criteria cited above.

We have also studied a few spectra that display only strong emission lines and have poor S/N, which have been classified 
as \ion{H}{II} regions (although the emission could actually be the result of the contribution from several \ion{H}{II} 
regions). Figure \ref{enana} illustrates this case, and  also includes an A-type star with low luminosity class and the 
white dwarf star. Stars with luminosity class V were discarded as members of NGC 55 and very likely belong to the 
Milky Way, since at the distance of NGC 55 a V-class star would be too faint to be observed. Although their radial 
velocities (see section 4) concur with the projected position on NGC 55, they are also consistent with the velocity of 
stars from the halo \citep{2006ApJ...647..303B}.

Twelve stars of the BA sample display the typical absorption lines of B- or A-type stars but broad emission in the wings 
of the Balmer lines. These might have a nebular origin, however the width of the wings suggests that they are dominated 
by stellar emission. A possible explanation for  these features is a strong wind, but we might be witnessing a Be star, 
electron scattering or even an outburst stage of LBV where other distinctive features (like emission in \ion{Fe}{II} lines) 
are masked by the low spectral resolution or are too weak (see \citealt{2000PASP..112...50W}, hereafter WF20). 
A\_42 (see Fig. \ref{Fig:wrtipo}) might be an extreme case,  where the Balmer lines are completely in emission (see below). 
We have marked these objects with {\it{'Str. winds?'} in Table \ref{t_catalog}}.

\subsection{{\bf{WRs \& LBV candidates}}}
\label{sec:wr}
There are 7 NGC~55 blue stars in the sample that do not belong to the normal O, B or A spectral bins. Most of these objects 
are candidate WR or LBV stars. We referred to "The OB Zoo" of WF20 and references therein for their  classification. The 
nitrogen WRs (WN) were also classified following the criteria of \cite{1996MNRAS.281..163S} based on nitrogen and helium 
lines. The classification of LBVs is particularly difficult, since they undergo different stages during their evolution 
\citep{2007AJ....134.2474M}. We looked for similarities with the spectra of known LBVs or LBV candidates (these objects 
are marked as LBVc in Table \ref{catalog1} and \ref{t_catalog}). The \ion{Fe}{II} emission lines were a decisive factor in 
our identification of these objects.

Figure~\ref{Fig:wrtipo} presents the spectra of these peculiar emission line stars found in the sample. C1\_18 shows the 
typical features of a WN8 star, namely intense emission lines of \ion{He}{II}$\lambda4686$ and \ion{N}{III}$\lambda4640$. 
The templates from WF20 render C1\_30 as an iron star and an LBV candidate. Its spectrum is rich in emission lines of  
hydrogen, helium and iron. Its overall morphology is very similar to that of the LBV ``$\eta$ Carinae", considered by 
WF20  as {{\it``the mother of all iron stars"}}. B\_13 displays H and $HeI$ P Cygni profiles with strong emission of the 
hydrogen lines, resembling the morphology of ``He3-519", a possible LBV candidate or WN11 according to WF20. B\_34 falls 
into the same category, although the P Cygni profiles are weaker. These stars present several similar features with a 
star studied by \cite{2002ApJ...577L.107B} in NGC 300, classified as Ofpe/WN9. C1\_51 is classified as an Ofpe/WN9 due 
to its similarity with the objects presented in this group by WF20 and the digital atlas of Ofpe/WN9 of 
\cite{1989PASP..101..520B}. B\_12 has typical features of Of stars, however we have classified this star as Of/WN due to 
the broad emission of \ion{He}{II}$\lambda4686$. The spectrum of A\_42 presents strong broad emission of Balmer lines that 
might be due to strong stellar winds. This is an A-star according to its \ion{Mg}{II}, \ion{Si}{II} and \ion{Ca}{II} 
lines, but in this case the spectrum shows transitions of \ion{Fe}{II} in emission. We conclude that this object is an 
LBV candidate. Finally, C1\_43 (not included in Fig. \ref{Fig:wrtipo}) shows features that might be interpreted as 
\ion{Fe}{II} in emission, but its S/N is too poor to be conclusive.

%%%%%%%%%%%%%%%%%%%%%%%%%%%%%%%%%%%%%%%%%%%%%%%%%%%%%%%%%%%%%%%%%%%%%%%%%%%%%%%%%%%%%%%%
%%%%%%%%%%%%%%%%%%%%%%%%%%%%%%%%%%%%%%%%%%%%%%%%%%%%%%%%%%%%%%%%%%%%%%%%%%%%%%%%%%%%%%%%

\section{RADIAL VELOCITY}

Radial velocities were determined from the Doppler wavelength shift of strong absorption lines (Balmer lines, \ion{He}{I}, 
\ion{He}{II} and metallic transitions of \ion{Mg}{II}, \ion{Si}{III}, depending on the spectral type). Alternative radial 
velocities were also evaluated from the emission lines of the surrounding nebula ([\ion{O}{III}]$\lambda5007$) when 
available. Radial velocities were calculated in the heliocentric system. For the dates of the observations the correction 
was $\sim$$-21$ km s$^{-1}$.

From the scatter in the individual determinations of the radial velocity of different lines we have obtained an average 
uncertainty of $\sim39$ km s$^{-1}$ (note that in the case of nebular emission we have only used one line therefore the 
accuracy is smaller but enough to compare with the stellar data). Additional error sources apply, such as the wavelength 
calibration, but are of second order by comparison.

\begin{figure}
 \resizebox{\hsize}{!}{\includegraphics{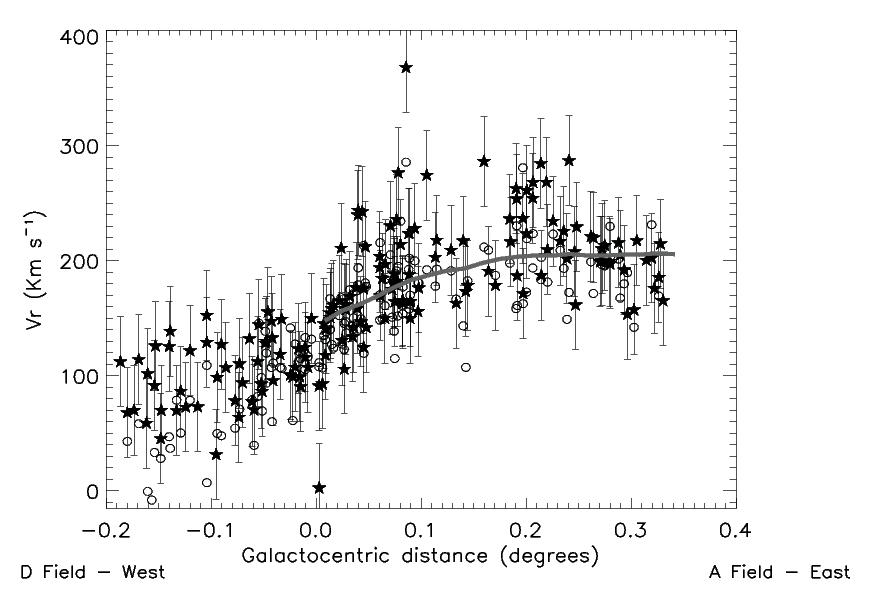}}	
    \caption{Stellar radial velocities of NGC 55 blue stars as a function of the projected galactocentric distance (stars)
    and radial velocities measured from nebular emission (open circles). The galaxy's rotational curve derived by 
    \cite{1991AJ....101..447P} from \ion{H}{I} regions (solid line) is included for comparison and there is good agreement 
    with our results.}
       \label{Fig:vra1}
 \end{figure}

The resulting radial velocities are plotted in Fig. \ref{Fig:vra1} as a function of the projected galactocentric distance, 
which is calculated with respect to the optical centre of NGC 55 ($\alpha$=$00h14m53.60s$  $\delta$=$-39d11m47.9s$  [J2000], 
\citealt{2003twoMASS}). The results from nebular  and stellar lines agree. We have also plotted the rotation curve 
derived from \ion{H}{I} observations by \cite{1991AJ....101..447P}. The track was shifted 1.5' to match the \ion{H}{I} 
calculated rotation centre and the optical centre. Figure~\ref{Fig:vra1} shows how the stellar population traces the 
rotation of the galaxy, even in its outer parts. We obtain a systematic velocity (i. e. the radial velocity of NGC 55) 
of $\sim$120 km s$^{-1}$, very similar to the 129 km s$^{-1}$ obtained by \cite{2004AJ....128...16K}. The fact that the 
young stellar population tracks the \ion{H}{I} rotation curve has also been found in NGC 3109 by \cite{2007ApJ...659.1198E}.

\begin{figure}
  \resizebox{\hsize}{!}{\includegraphics{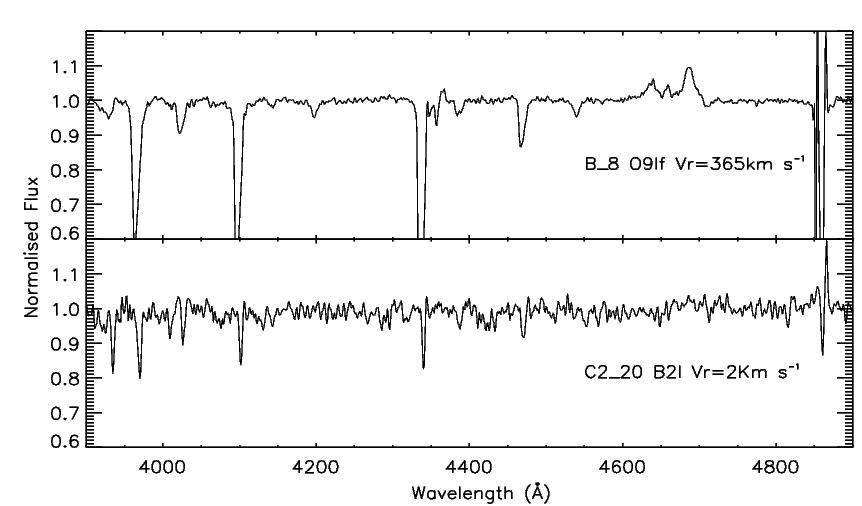}}	
    \caption{Spectra of objects with anomalous radial velocity.}
       \label{Fig:vra2}
 \end{figure}

The error bars of the results and the high inclination angle of NGC 55 prevent us from studying substructures in the 
galaxy, even though a few objects exhibit a radial velocity far from the normal behaviour of the stars. C2\_20 (B2I) has a 
very small radial velocity while B\_8 (O9If) presents a high radial velocity for its position in the galaxy. Figure 
\ref{Fig:vra2} shows the spectra of these stars.

\begin{figure}
  \resizebox{\hsize}{!}{\includegraphics{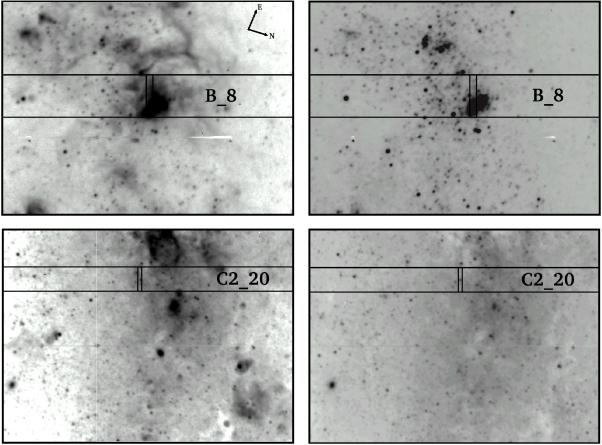}}
    \caption{V-band (right) and H${\alpha}$ (left) images taken with FORS2. The position of the slits used to observe 
    B\_8 (top) and C2\_20 (bottom) are marked on the image. B\_8 is near large \ion{H}{II} structures, while C2\_20 is 
    located close the centre of NGC55.} 
       \label{Fig:vraHa}
 \end{figure}

B\_8 is located in the middle of a cluster with \ion{H}{II} regions seen in emission in the H${\alpha}$ image (see 
Fig. \ref{Fig:vraHa}). Although this object presents absorption lines typical of an O-star, we are probably observing a 
group of stars. Its bright magnitude compared to other O-stars and the shape of the line profiles agree with this 
hypothesis. In addition, the spectrum is clearly affected by the nebular emission around the object.

The spectrum of C2\_20 exhibits no hint as to the reason for its small radial velocity. We have estimated a lower 
metallicity for C2\_20 (see section 5) and given its high galactic latitude, it might be a Galactic star of the halo with 
low metal content. However, this would be completely inconsistent with its classification as a supergiant since it would 
be brighter. Moreover, it is in the centre of NGC 55 and close to different clusters which show evidence of intense stellar 
activity. Therefore, we cannot exclude the possibilities that either a supernova explosion or binarity can be the cause of 
its peculiar radial velocity, although we must be cautious with this object due to the moderate S/N of its spectrum.

%%%%%%%%%%%%%%%%%%%%%%%%%%%%%%%%%%%%%%%%%%%%%%%%%%%%%%%%%%%%%%%%%%%%%%%%%%%%%%%%%%%%%%%%
%%%%%%%%%%%%%%%%%%%%%%%%%%%%%%%%%%%%%%%%%%%%%%%%%%%%%%%%%%%%%%%%%%%%%%%%%%%%%%%%%%%%%%%%

\section{METALLICITY: A ROUGH ESTIMATE}

Studies based on different methods (\cite{2003A&A...412...69T}, \ion{H}{II} regions; \cite{2005ApJ...622..279D}, stellar 
content) show that the chemical composition in the plane of NGC 55 is close to that of the LMC. We present here a 
qualitative estimate of the metal content of NGC 55. A subset of spectral lines of our stars are compared with objects of 
similar spectral type from the Milky Way (MW) \citep{1992A&AS...94..569L}, LMC \citep{2006A&A...456..623E}\footnote{These 
spectra were obtained in the VLT-FLAMES survey of massive stars in the Milky Way and the Magellanic Clouds 
(http://star.pst.qub.ac.uk/$\sim$sjs/flames/).} and SMC \citep{1997A&A...317..871L}. This method has several caveats; 
the quality of our data, the use of different atlases for contrast and possible variations of the chemical composition of 
the comparison objects. While we have corrected for the diverse resolution of the standard spectra, we cannot account for 
other factors such as data reduction methods or slight differences in the spectral classification outcome. Nonetheless, 
the results will give us a flavour of the mean metallicity of the young stellar population and the possible presence of a 
chemical gradient. In a forthcoming paper we will perform a quantitative abundance analysis of a subset of blue massive 
stars in NGC 55.

The metallicity is estimated by comparing the strength of the lines of silicon (\ion{Si}{IV}$\lambda4089$, 
\ion{Si}{III}$\lambda4552-4568-4575$, \ion{Si}{II}$\lambda4128-4130$) and magnesium (\ion{Mg}{II}$\lambda4481$) found in 
NGC 55 stars to those observed in MW, SMC and LMC objects. Because the strength of these metallic lines depends on the 
stellar effective temperature and gravity, we have always used stars of the same spectral type and luminosity class (or as 
close as possible) for the comparison. Strong photospheric lines of carbon, oxygen and nitrogen exist in this wavelength 
range, but they can be contaminated by processed material from deeper layers due to mixing and evolution 
\citep{2000ARA&A..38..143M,2008ApJ...676L..29H}. Additional studies report a non-negligible underabundance of nitrogen in 
NGC 55 (possibly due to the galaxy being either young or deficient in lower mass stars relative to the LMC, 
\citealt{1983MNRAS.204..743W,1985Ap&SS.117..271A}).

\begin{figure}
    \resizebox{\hsize}{!}{\includegraphics[angle=0]{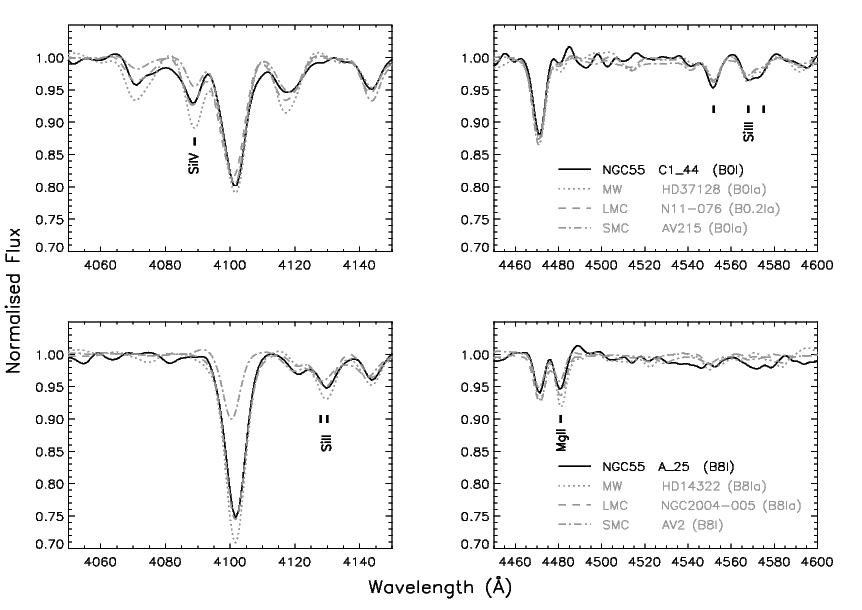}}
      \caption{Example of  metallicity determination: the spectra of two NGC 55 stars, C1\_44 (top) and A\_25 (bottom), 
      are compared to objects from the Milky Way, LMC and SMC with similar spectral types. In these cases the strength 
      of the silicon and magnesium lines of the NGC 55 stars is similar to the LMC stars.}
         \label{Fig:metalcomp}
   \end{figure}

To illustrate the method two sample stars are compared to MW and Magellanic Clouds stars of similar spectral type in 
Fig. \ref{Fig:metalcomp}. The strength of  silicon and magnesium transitions in the NGC 55 stars of the example resemble 
that of their LMC counterparts. The same analysis was performed on a set of supergiant stars between B0 and A0 
(77 objects). We have distributed the stars into five metallicity bins (MW, MW--LMC, LMC, LMC--SMC, SMC) as we 
indicate in Tables \ref{catalog1} and \ref{t_catalog}.

\begin{figure}
   \resizebox{\hsize}{!}{\includegraphics[angle=0]{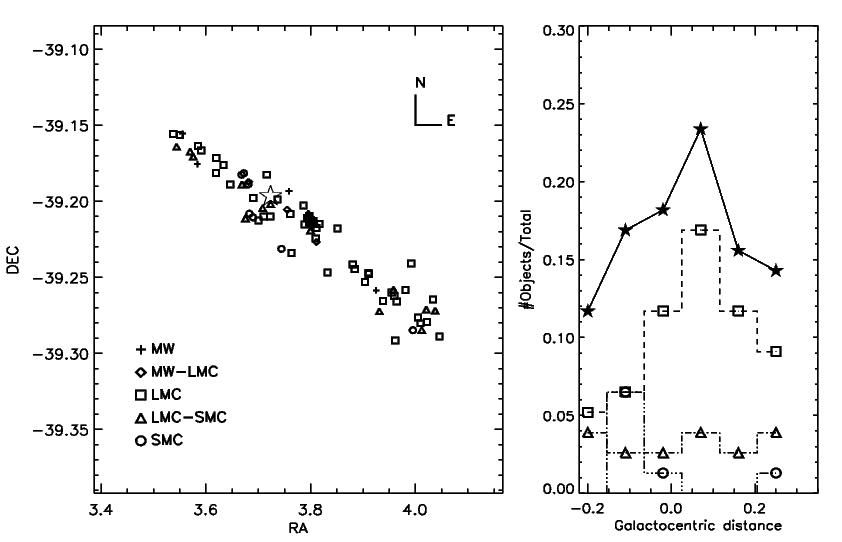}}
     \caption{Metallicity distribution of B0-A0 supergiants in NGC 55.  Left: location of the sample stars in the galaxy,
      the different symbols indicate their metallicity as shown in the plot. The star shows the optical centre 
      of the galaxy. The average metallicity is close to the LMC. Right: The number of objects per metallicity bin 
      normalised by the total of stars considered, as a function of projected galactocentric radio (degrees). 
      The histogram hints a variation across the projected disk of NGC 55. The solid line represents the stars in each 
      bin over the total, thus giving an idea of how large is the sample in each position bin. We have not plotted the 
      objects with MW and LMC-MW metallicity for clarity, since there are only four of them.}
	\label{Fig:metaldis}
  \end{figure}

We find that the average metal content of the NGC 55 supergiants is similar to those of the LMC. At first glance the 
spatial distribution (see Fig. \ref{Fig:metaldis}, left) does not show any systematic variation across the galaxy, in 
agreement with \cite{1983MNRAS.204..743W} who determined oxygen abundances  from a small sample of bright \ion{H}{II} 
regions in NGC 55. Nevertheless, if we plot the number of objects normalised by the total number of stars as a function 
of the projected galactocentric distance (see Fig. \ref{Fig:metaldis}, right), some variation can be observed. At a 
projected galactocentric distance of 0.15 degrees and greater the population of stars with  lower metallicity increases 
compared to the central population, where the majority of stars have LMC metallicity.

Once again, the high inclination angle of NGC 55 complicates our study. We cannot ensure that stars apparently close to 
the centre are indeed in the disk of the galaxy and not in an outer layer but in the line of sight towards the centre. 
While the indirect evidence of a radial gradient found here cannot be considered significant until we perform a 
quantitative analysis, our exercise excludes a peculiar behaviour in our sample.

%%%%%%%%%%%%%%%%%%%%%%%%%%%%%%%%%%%%%%%%%%%%%%%%%%%%%%%%%%%%%%%%%%%%%%%%%%%%%%%%%%%%%%%%
%%%%%%%%%%%%%%%%%%%%%%%%%%%%%%%%%%%%%%%%%%%%%%%%%%%%%%%%%%%%%%%%%%%%%%%%%%%%%%%%%%%%%%%%

\section{SUMMARY AND CONCLUSIONS}

The long-term goal of our project is to obtain the physical parameters and abundances of blue massive stars in NGC 55. 
Results will be useful to study the evolution of massive stars and its dependence with metallicity. The abundance 
measurements, representative of the clouds where the stars were born, are valuable input parameters to study the chemical 
evolution of the host galaxy. Another priority goal is to apply and test the Flux-Weighted Gravity-Luminosity Relation 
\citep{2008ApJ...999...999} to the blue supergiant stars in NGC 55 to measure an independent spectroscopic distance to 
the galaxy. NGC 55 will be an excellent target for this given its abundant blue supergiant population.

As a first step, we have presented in this paper the first census of blue massive stars in NGC 55. The stars were 
identified from a qualitative analysis of their low-resolution optical spectra that yielded spectral classification 
for 204 objects with blue colours, out of which 164 have types earlier than F0.

We have also determined radial velocities from the observed spectra. In spite of the low wavelength accuracy and spectral 
resolution, we see a good agreement with the published rotation curve measured from \ion{H}{I} 
\citep{1991AJ....101..447P}. This study allowed us to detect one object with peculiar velocity that deserves a more 
detailed study.

A preliminary estimate of stellar metallicity, from the comparison of spectra with stars in the Milky Way, LMC and SMC, 
shows that the mean metal content of the stars is close to that of the LMC with some indication of  a variation along the 
projected galactocentric distance. This requires confirmation  by quantitative studies.

In a forthcoming paper, we plan to analyse the spectra of those blue massive stars with the highest S/N spectra with 
FASTWIND (Fast Analysis of STellar atmospheres with WINDs, \citealt{1997A&A...323..488S,2005A&A...435..669P}).

%%%%%%%%%%%%%%%%%%%%%%%%%%%%%%%%%%%%%%%%%%%%%%%%%%%%%%%%%%%%%%%%%%%%%%%%%%%%%%%%%%%%%%
%%%%%%%%%%%%%%%%%%%%%%%%%%%%%%%%%%%%%%%%%%%%%%%%%%%%%%%%%%%%%%%%%%%%%%%%%%%%%%%%%%%%%%

\begin{acknowledgements}

The authors would like to thank  D. J. Lennon, M. A. Urbaneja and C. Evans for their help with the spectral classification 
and useful comments.  D. J. Lennon is also acknowledged for his careful reading of the manuscript. We thank the 
FLAMES Survey of Massive Stars project (P.I. S. Smartt) for the LMC spectra. Finally, we are thankful to the referee for 
his/hers comments. WG and GP gratefully acknowledge financial support for this work from the Chilean Center for Astrophysics 
FONDAP 15010003. This project has been supported by Spanish grants AYA2004-08271-CO2-O1 and AYA2007-67456-C02-01.

\end{acknowledgements}

%############################################################
%	BIBLIOGRAPHY
%############################################################

\bibliographystyle{aa}

\bibliography{9399Bibl}
\clearpage
\begin{appendix}
\section{Catalogue}
\input{9399ApA1}
\input{9399ApA2}
\input{9399ApA3}
\section{Atlas}
\input{9399ApB1}

%%%%%%%%%%%%%%%%%%% FINDING CHART %%%%%%%%%%%%%%%%%%%%%%%%%%%%%%%%%%%%%%%%%%%%%%%%%%

\clearpage
\section{Finding Charts}
   \begin{figure*}[!h]
   \centering
   \includegraphics[angle=0,width=15cm]{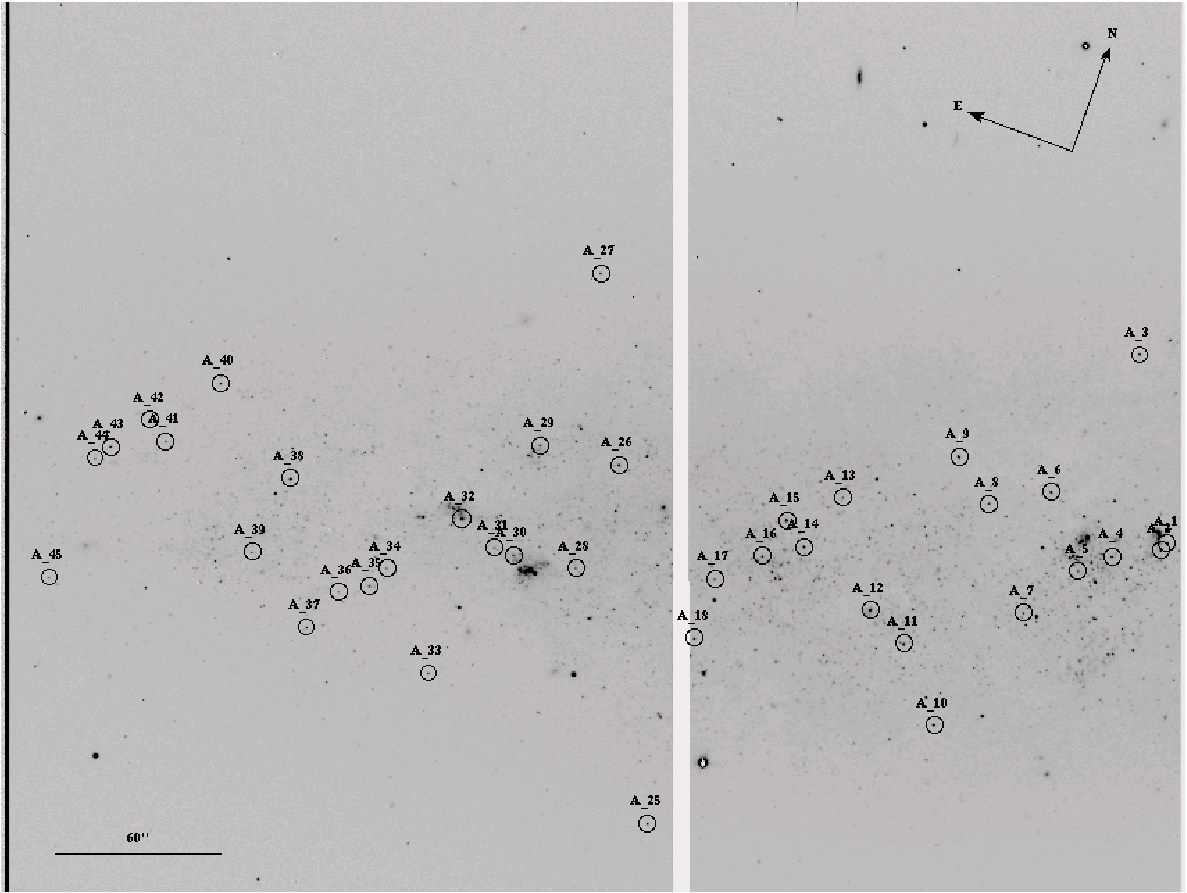}
      \caption{FORS2 V-band preimaging of field A. The stars studied in this work are labelled and encircled.}
       \label{FindA}
   \end{figure*}

   \begin{figure*}[!h]
   \centering
   \includegraphics[angle=0,width=15cm]{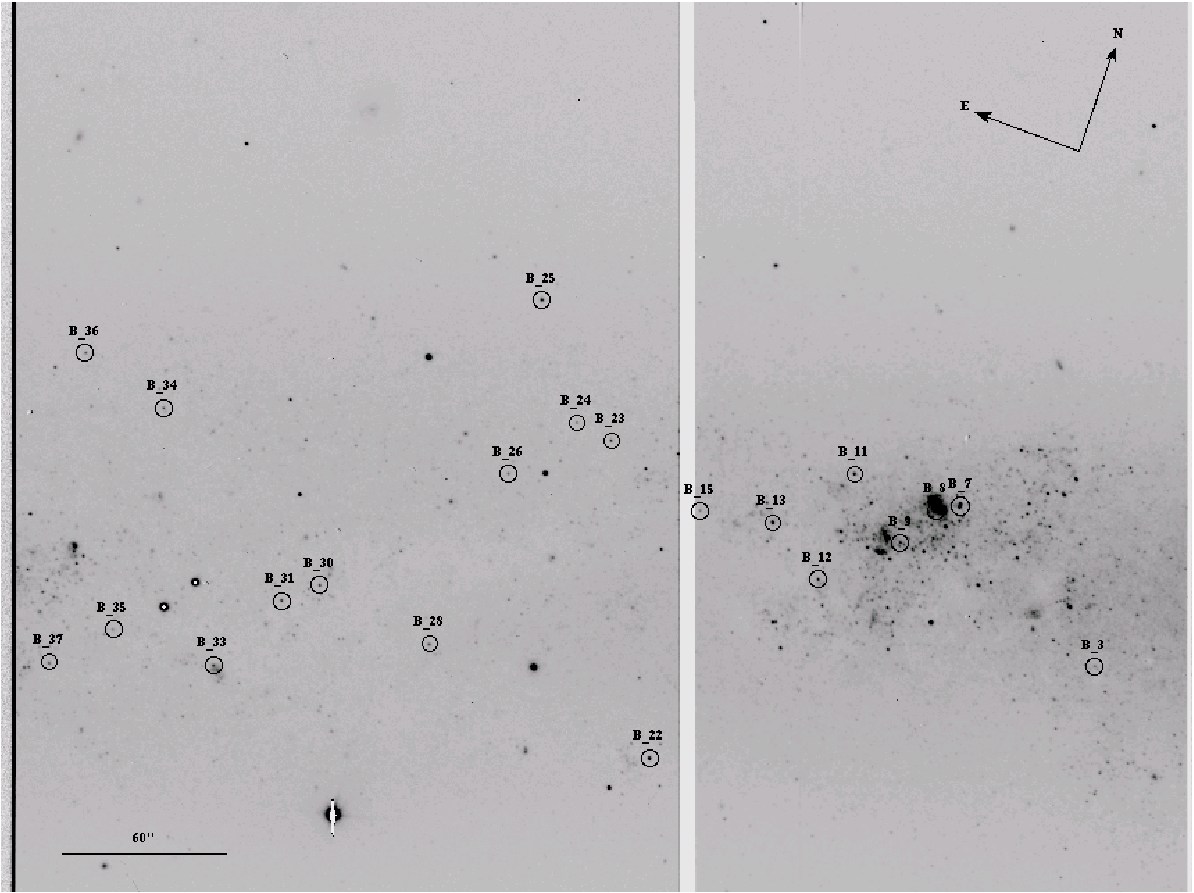}
      \caption{Same as Fig. \ref{FindA}. Target stars in field B.}
         \label{FindB}
   \end{figure*}

\clearpage

   \begin{figure*}
   \centering
   \includegraphics[angle=0,width=15cm]{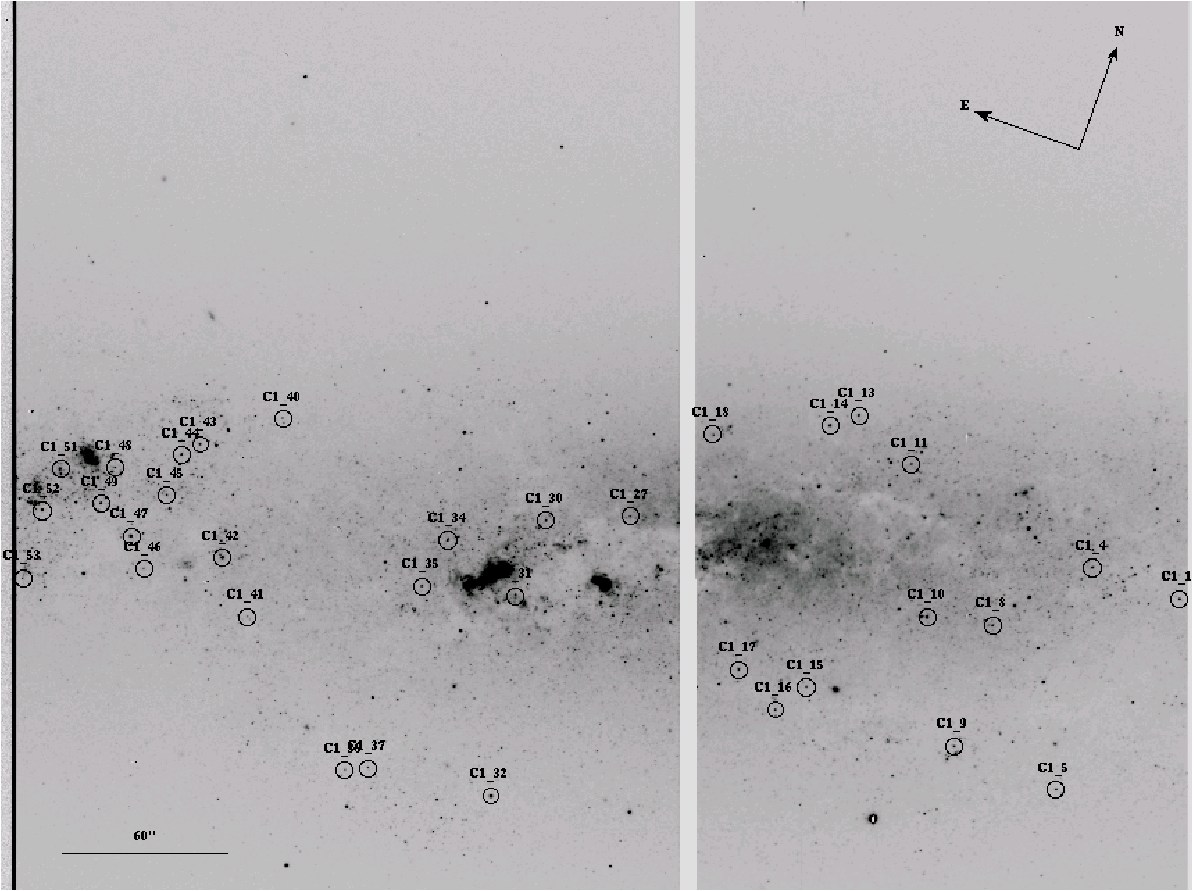}
      \caption{Same as Fig. \ref{FindA}. Target stars in field C observed through mask C1.}
         \label{FindC1}
   \end{figure*}
   
   \begin{figure*}
   \centering
   \includegraphics[angle=0,width=15cm]{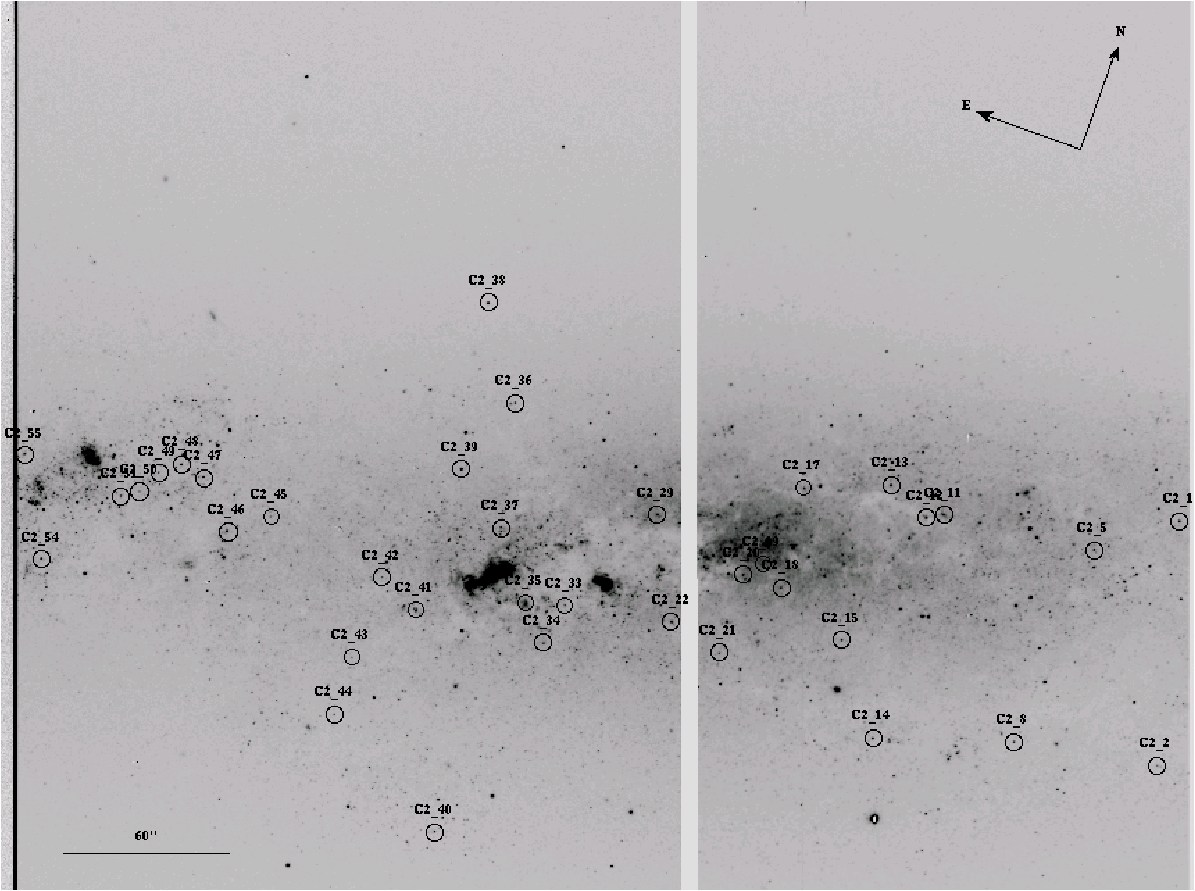}
      \caption{Same as Fig. \ref{FindA}. Target stars in field C observed through mask C2.}
         \label{FindC2}
   \end{figure*}

\clearpage
   \begin{figure*}
   \centering
   \includegraphics[angle=0,width=15cm]{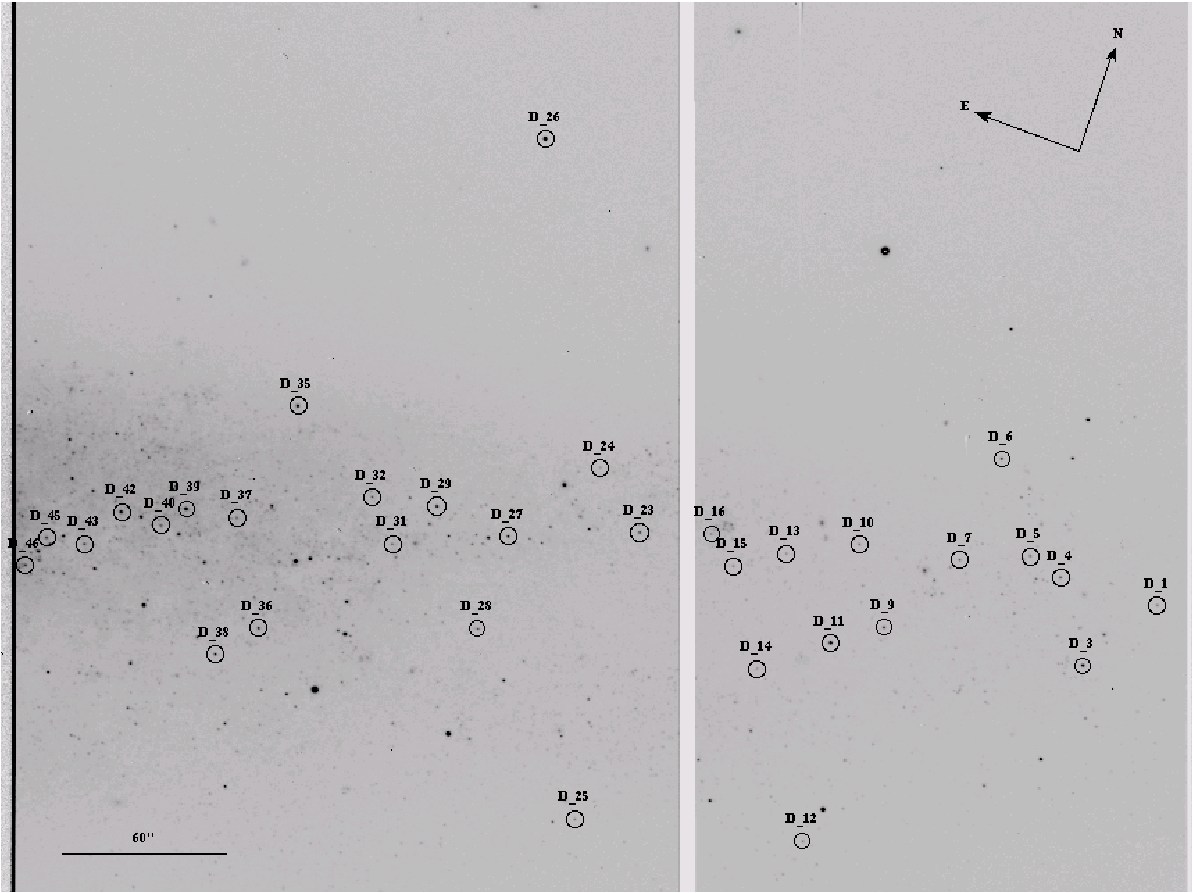}
      \caption{Same as Fig. \ref{FindA}. Target stars in field D.}
         \label{FindD}
   \end{figure*}

%%%%%%%%%%%%%%%%%%%%%%%%%%%%%%%%%%%%%%%%%%%%%%%%%%%%%%%%%%%%%%%%%%%%%
\end{appendix}
\end{document}

%% file: 9399ApA1.tex
\longtab{1}{
\begin{longtable}{ccccclccccl}
\caption{\label{t_catalog} Blue massive stars observed in NGC 55. We have also included spectra displaying only  
 \ion{H}{II} emissions. The columns list: (1) star identification, consisting on
 field number followed by slit number; (2) right ascension (hh:mm:ss); (3) declination (dd:mm:ss); (4) V-magnitude; 
(5) V-I colour; (6) spectral type; (7) projected galactocentric distance (arc-seconds); (8) radial velocity (km s$^{-1}$); 
(9) signal-to-noise ratio of the spectrum; (10) metallicity estimate; (11) comments.} \\
\hline\hline
\hline
\endfirsthead
 \footnotesize{ID}  & \footnotesize{ RA(J2000)} &  \footnotesize{DEC(J2000)}  &  \footnotesize{V}  &  \footnotesize{V-I}  & \footnotesize{Spectral Type} & \footnotesize{Rg} & \footnotesize{$v_r$}  &  \footnotesize{S/N}   &  \footnotesize{Metallicity}  &  \footnotesize{Comments}  \\
 \footnotesize{(1)}  & \footnotesize{(2)} &  \footnotesize{(3)}  &  \footnotesize{(4)}  &  \footnotesize{(5)}  & \footnotesize{(6)} & \footnotesize{(7)} & \footnotesize{(8)} &  \footnotesize{(9)}  &  \footnotesize{(10)}  &  \footnotesize{(11)}  \\
\hline
 \footnotesize{    D\_1}  &  \footnotesize{  0:14:08.97}  &  \footnotesize{-39:09:20.88}  &  \footnotesize{20.212}  &  \footnotesize{-0.557}  &  \footnotesize{        B9I}  &  \footnotesize{ -671.79}  &  \footnotesize{ 112}  &  \footnotesize{  43}  &  \footnotesize{     LMC}  &  \footnotesize{  		 }  \\
 \footnotesize{    D\_3}  &  \footnotesize{  0:14:10.51}  &  \footnotesize{-39:09:50.76}  &  \footnotesize{18.581}  &  \footnotesize{-0.407}  &  \footnotesize{        A0I}  &  \footnotesize{ -647.94}  &  \footnotesize{  67}  &  \footnotesize{ 117}  &  \footnotesize{ LMC-SMC}  &  \footnotesize{        Str. winds?}  \\
 \footnotesize{    D\_4}  &  \footnotesize{  0:14:12.11}  &  \footnotesize{-39:09:23.41}  &  \footnotesize{19.724}  &  \footnotesize{-0.524}  &  \footnotesize{        B9I}  &  \footnotesize{ -625.02}  &  \footnotesize{  69}  &  \footnotesize{  60}  &  \footnotesize{     LMC}  &  \footnotesize{  		 }  \\
 \footnotesize{    D\_5}  &  \footnotesize{  0:14:13.22}  &  \footnotesize{-39:09:19.79}  &  \footnotesize{19.963}  &  \footnotesize{-0.813}  &  \footnotesize{        B3I}  &  \footnotesize{ -608.56}  &  \footnotesize{ 113}  &  \footnotesize{  44}  &  \footnotesize{	MW}  &  \footnotesize{  		 }  \\
\hline
\hline
\end{longtable}
}% End \longtab

%% file: 9399ApA2.tex
\longtab{2}{
\begin{longtable}{ccccclccccl}
\caption{\label{t_catalogF} F- and G-type stars observed in NGC 55. Columns are detailed in Table \ref{t_catalog}.} \\
\hline\hline
\hline
\endfirsthead
 \footnotesize{ID}  & \footnotesize{ RA(J2000)} &  \footnotesize{DEC(J2000)}  &  \footnotesize{V}  &  \footnotesize{V-I}  & \footnotesize{Spectral Type} & \footnotesize{Rg} & \footnotesize{$v_r$}  &  \footnotesize{S/N}   &  \footnotesize{Metallicity}  &  \footnotesize{Comments}  \\
 \footnotesize{(1)}  & \footnotesize{(2)} &  \footnotesize{(3)}  &  \footnotesize{(4)}  &  \footnotesize{(5)}  & \footnotesize{(6)} & \footnotesize{(7)} & \footnotesize{(8)} &  \footnotesize{(9)}  &  \footnotesize{(10)}  &  \footnotesize{(11)}  \\
\hline
  \footnotesize{    D\_2}  &   \footnotesize{  0:14:11.65}  &   \footnotesize{-39:08:19.69}  &   \footnotesize{20.461}  &   \footnotesize{-0.213}  &   \footnotesize{               F0-2I}  &   \footnotesize{ -634.56}  &   \footnotesize{  17}  &   \footnotesize{  36}  &   \footnotesize{       -}  &   \footnotesize{                    }  \\
  \footnotesize{   D\_33}  &   \footnotesize{  0:14:32.48}  &   \footnotesize{-39:11:11.40}  &   \footnotesize{17.808}  &   \footnotesize{-0.223}  &   \footnotesize{                 F5I}  &   \footnotesize{ -317.28}  &   \footnotesize{  56}  &   \footnotesize{ 166}  &   \footnotesize{       -}  &   \footnotesize{                 neb.}  \\
  \footnotesize{   D\_34}  &   \footnotesize{  0:14:33.65}  &   \footnotesize{-39:10:56.64}  &   \footnotesize{19.475}  &   \footnotesize{-0.051}  &   \footnotesize{                 F5I}  &   \footnotesize{ -300.21}  &   \footnotesize{  30}  &   \footnotesize{  46}  &   \footnotesize{       -}  &   \footnotesize{                 neb.}  \\
  \footnotesize{   D\_41}  &   \footnotesize{  0:14:39.70}  &   \footnotesize{-39:10:56.64}  &   \footnotesize{18.418}  &   \footnotesize{-0.024}  &   \footnotesize{               F5-8I}  &   \footnotesize{ -210.06}  &   \footnotesize{  55}  &   \footnotesize{  76}  &   \footnotesize{       -}  &   \footnotesize{                    }  \\
\hline
\hline
\end{longtable}
}% End \longtab

%% file: 9399ApA3.tex
\longtab{3}{
\begin{longtable}{ccccclccccl}
\caption{\label{t_catalogMW} Low luminosity class stars, likely members of the Milky Way halo. Columns 
 are the same as  in Table \ref{t_catalog}.} \\
\hline\hline
\hline
\endfirsthead
 \footnotesize{ID}  & \footnotesize{ RA(J2000)} &  \footnotesize{DEC(J2000)}  &  \footnotesize{V}  &  \footnotesize{V-I}  & \footnotesize{Spectral Type} & \footnotesize{Rg} & \footnotesize{$v_r$}  &  \footnotesize{S/N}   &  \footnotesize{Metallicity}  &  \footnotesize{Comments}  \\
 \footnotesize{(1)}  & \footnotesize{(2)} &  \footnotesize{(3)}  &  \footnotesize{(4)}  &  \footnotesize{(5)}  & \footnotesize{(6)} & \footnotesize{(7)} & \footnotesize{(8)} &  \footnotesize{(9)}  &  \footnotesize{(10)}  &  \footnotesize{(11)}  \\
\hline
 \footnotesize{    D\_8}  &  \footnotesize{  0:14:16.16}  &  \footnotesize{-39:09:38.16}  &  \footnotesize{18.945}  & \footnotesize{-0.161}  &  \footnotesize{  	       A5V}  &  \footnotesize{ -564.09}  &  \footnotesize{  75}  &  \footnotesize{  69}  &  \footnotesize{	 -}  &  \footnotesize{  	       neb.}  \\
 \footnotesize{   D\_30}  &  \footnotesize{  0:14:32.02}  &  \footnotesize{-39:10:17.03}  &  \footnotesize{19.746}  & \footnotesize{ 0.139}  &  \footnotesize{  	     G2-5V}  &  \footnotesize{ -326.59}  &  \footnotesize{  34}  &  \footnotesize{  36}  &  \footnotesize{	 -}  &  \footnotesize{  		  }  \\
 \footnotesize{   C1\_3}  &  \footnotesize{  0:14:40.77}  &  \footnotesize{-39:12:01.07}  &  \footnotesize{19.016}  & \footnotesize{-1.017}  &  \footnotesize{  		Da}  &  \footnotesize{ -192.58}  &  \footnotesize{ 106}  &  \footnotesize{ 111}  &  \footnotesize{	 -}  &  \footnotesize{  		  }  \\
 \footnotesize{   C2\_3}  &  \footnotesize{  0:14:42.07}  &  \footnotesize{-39:11:17.16}  &  \footnotesize{19.045}  & \footnotesize{-0.020}  &  \footnotesize{  	       B2V}  &  \footnotesize{ -173.70}  &  \footnotesize{ 107}  &  \footnotesize{  40}  &  \footnotesize{	 -}  &  \footnotesize{  		  }  \\
\hline
\hline
\end{longtable}
}% End \longtab

%% file: 9399ApB1.tex
\clearpage
\begin{figure*}
\includegraphics[width=18cm]{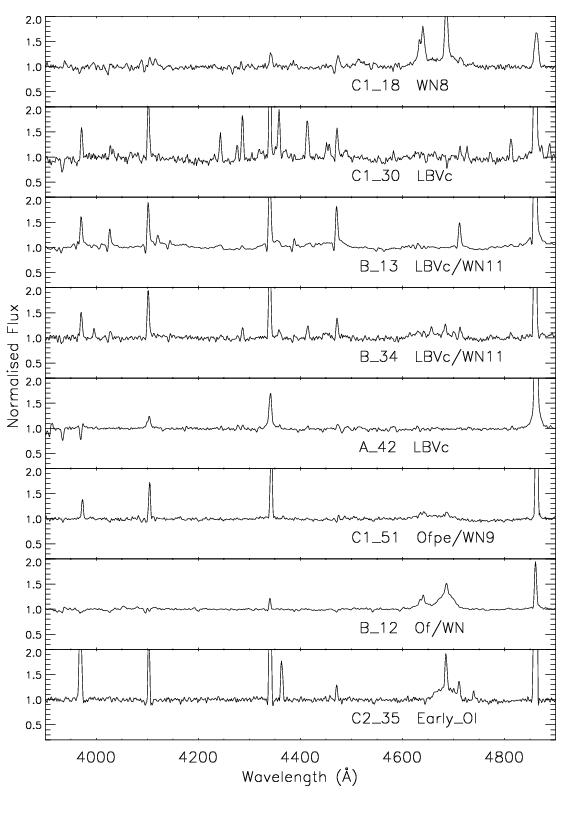}
\caption{Digital spectral atlas of VLT-FORS2 observations of NGC 55. LBV candidates (LBVc), WRs
and Of-stars.}
\label{ape0}
\end{figure*}

\clearpage
\begin{figure*}
\includegraphics[width=18cm]{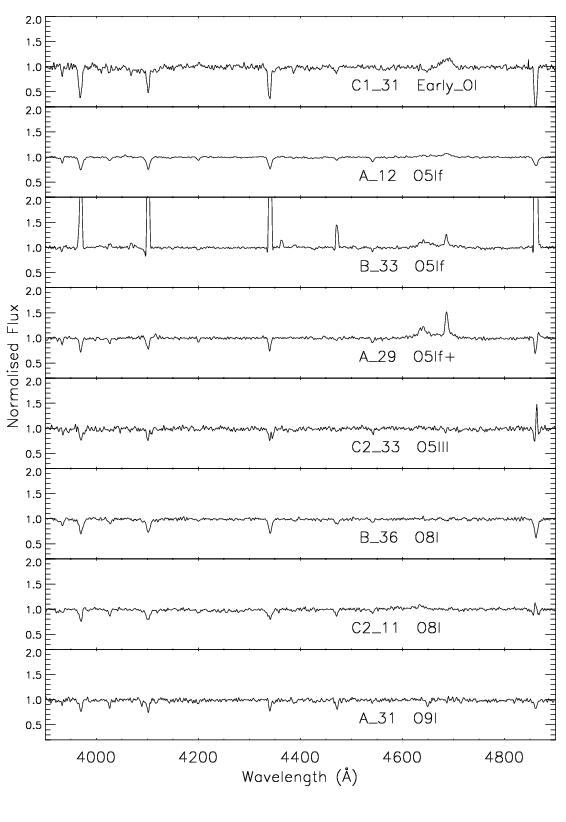}
\caption{O-type supergiants in NGC 55.}
\label{ape1}
\end{figure*}

\clearpage
\begin{figure*}
\includegraphics[width=18cm]{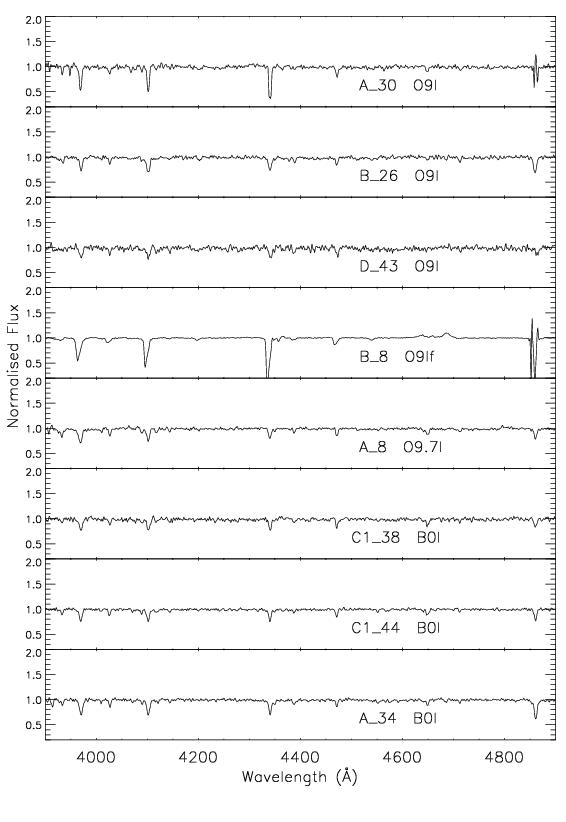}
\caption{Late-O and B0 supergiants in NGC 55.}
\label{ape2}
\end{figure*}

\clearpage
\begin{figure*}
\includegraphics[width=18cm]{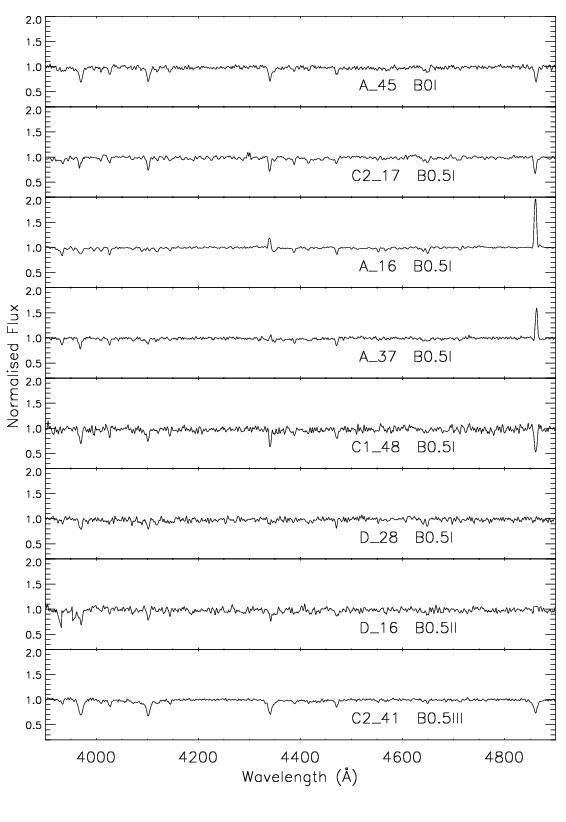}
\caption{B0 stars in NGC 55.}
\label{ape3}
\end{figure*}

\clearpage
\begin{figure*}
\includegraphics[width=18cm]{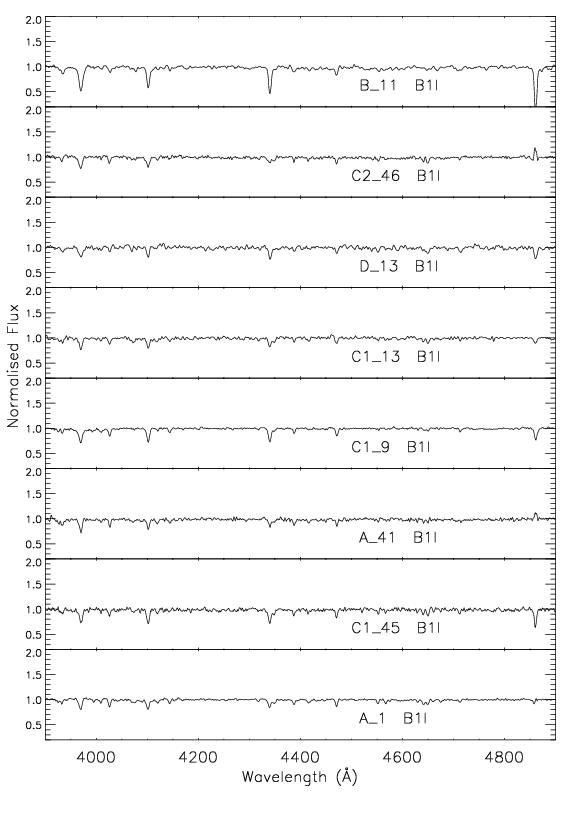}
\caption{B1 supergiants in NGC 55.}
\label{ape4}
\end{figure*}

\clearpage
\begin{figure*}
\includegraphics[width=18cm]{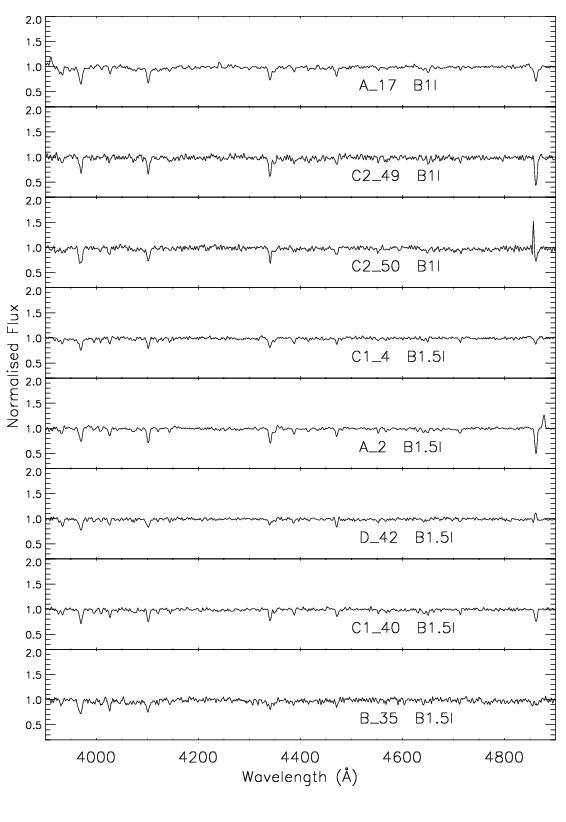}
\caption{B1 supergiants in NGC 55, continued.}
\label{ape5}
\end{figure*}

\clearpage
\begin{figure*}
\includegraphics[width=18cm]{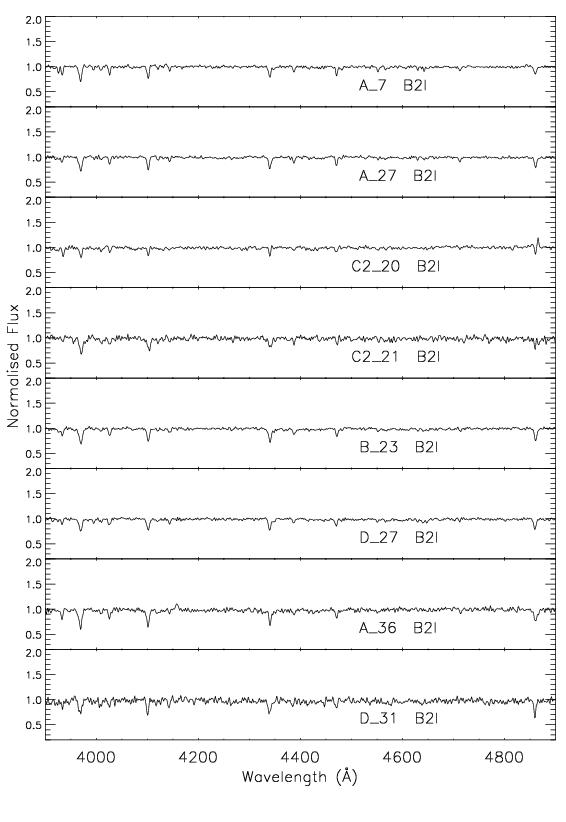}
\caption{B2 supergiants in NGC 55.}
\label{ape6}
\end{figure*}

\clearpage
\begin{figure*}
\includegraphics[width=18cm]{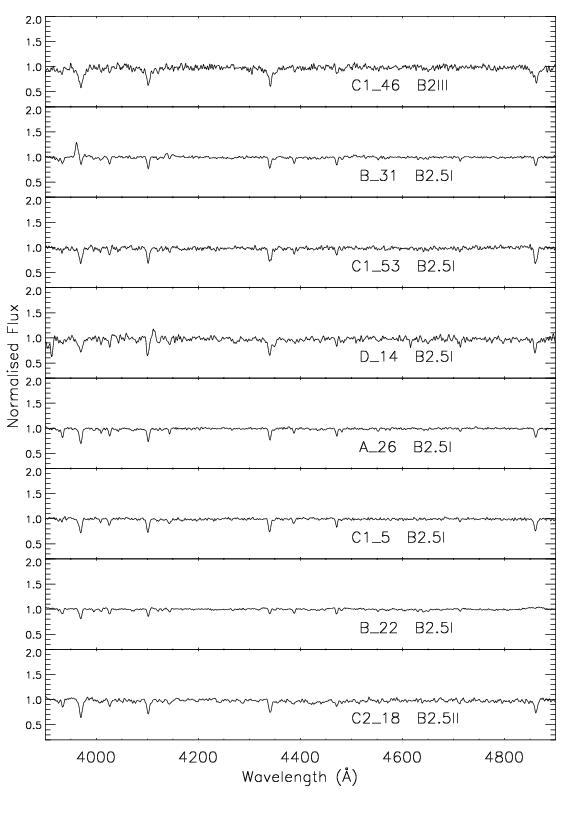}
\caption{B2 supergiants in NGC 55, continued.}
\label{ape7}
\end{figure*}

\clearpage
\begin{figure*}
\includegraphics[width=18cm]{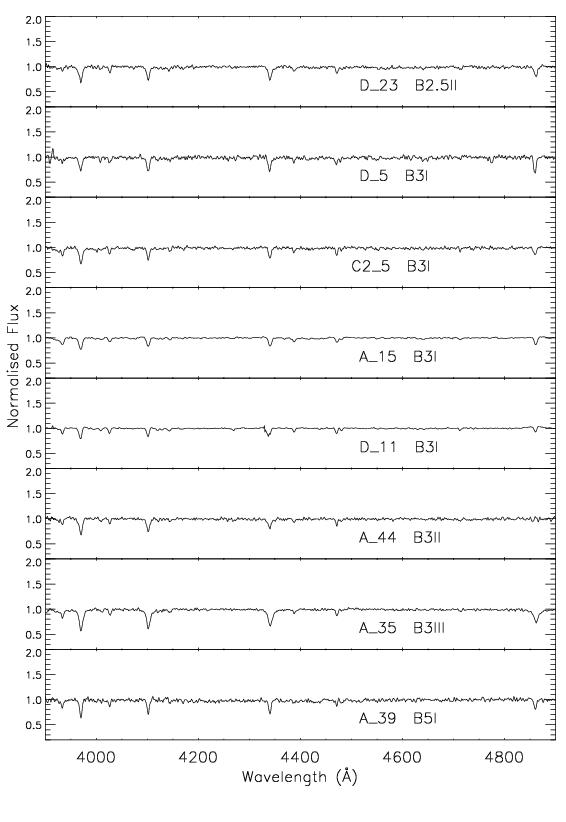}
\caption{Mid-B stars in NGC 55.}
\label{ape8}
\end{figure*}

\clearpage
\begin{figure*}
\includegraphics[width=18cm]{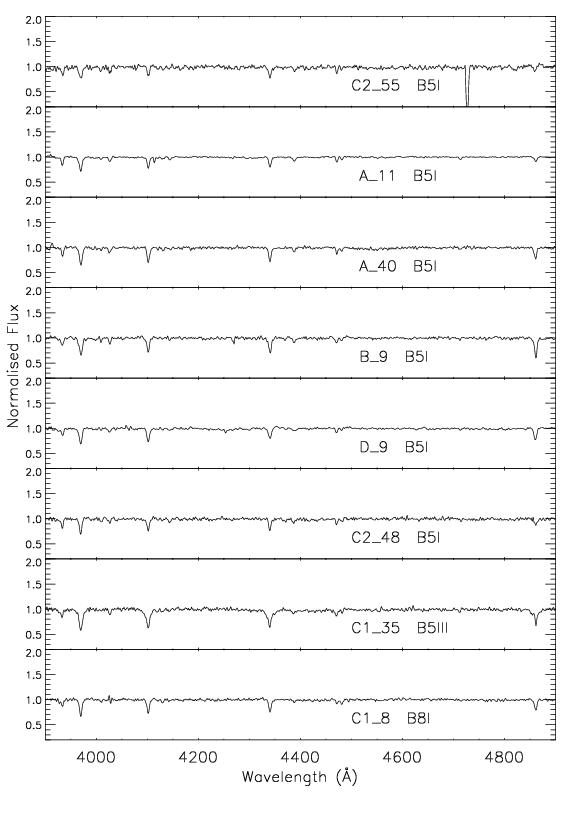}
\caption{Mid-B stars in NGC 55, continued.}
\label{ape9}
\end{figure*}

\clearpage
\begin{figure*}
\includegraphics[width=18cm]{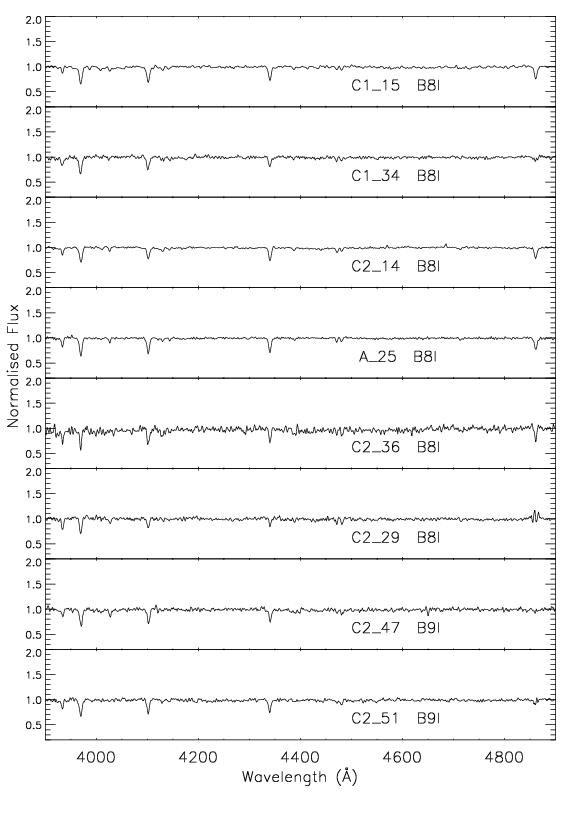}
\caption{Late-B supergiants in NGC 55.}
\label{ape10}
\end{figure*}

\clearpage
\begin{figure*}
\includegraphics[width=18cm]{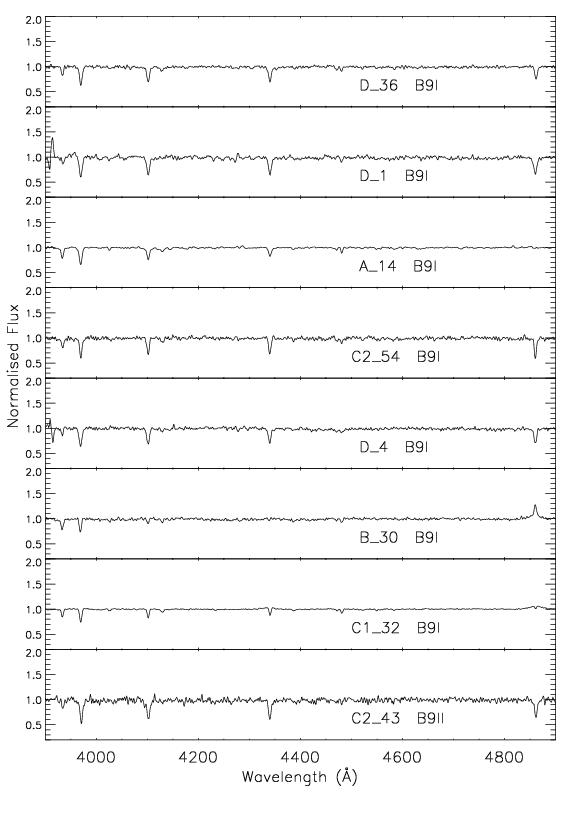}
\caption{Late-B supergiants in NGC 55, continued.}
\label{ape11}
\end{figure*}

\clearpage
\begin{figure*}
\includegraphics[width=18cm]{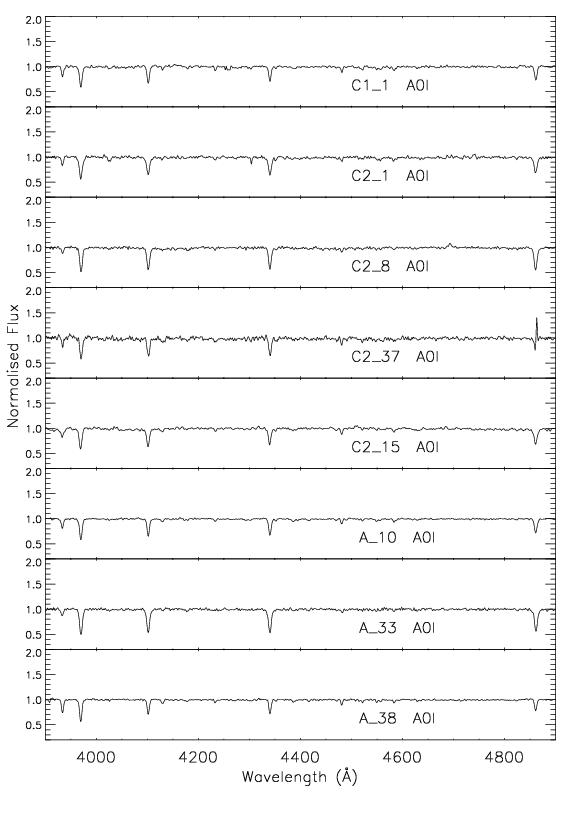}
\caption{A0 supergiants in NGC 55.}
\label{ape12}
\end{figure*}

\clearpage
\begin{figure*}
\includegraphics[width=18cm]{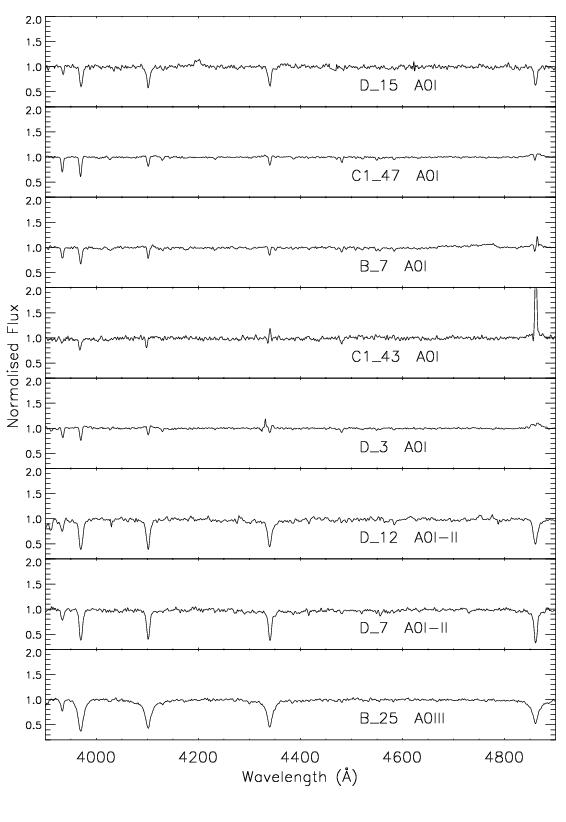}
\caption{A0 stars in NGC 55, continued.}
\label{ape13}
\end{figure*}

\clearpage
\begin{figure*}
\includegraphics[width=18cm]{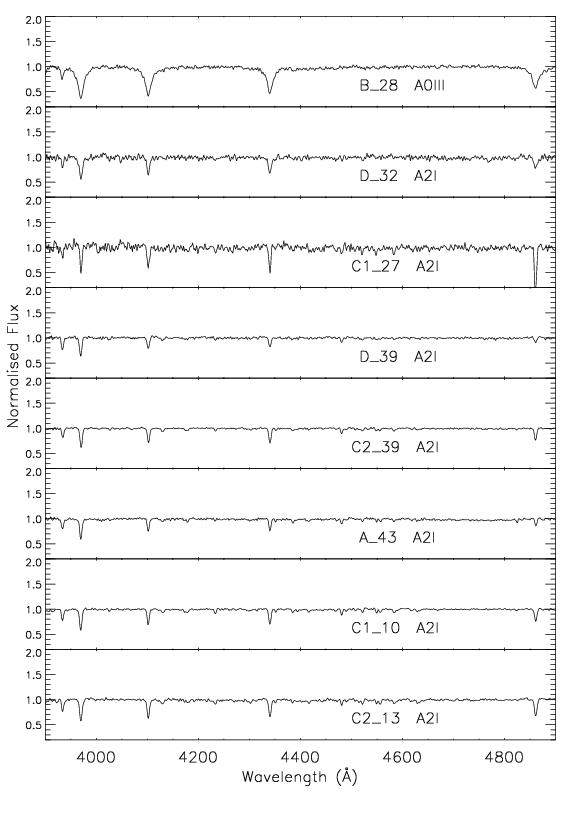}
\caption{Early-A stars in NGC 55.}
\label{ape14}
\end{figure*}

\clearpage
\begin{figure*}
\includegraphics[width=18cm]{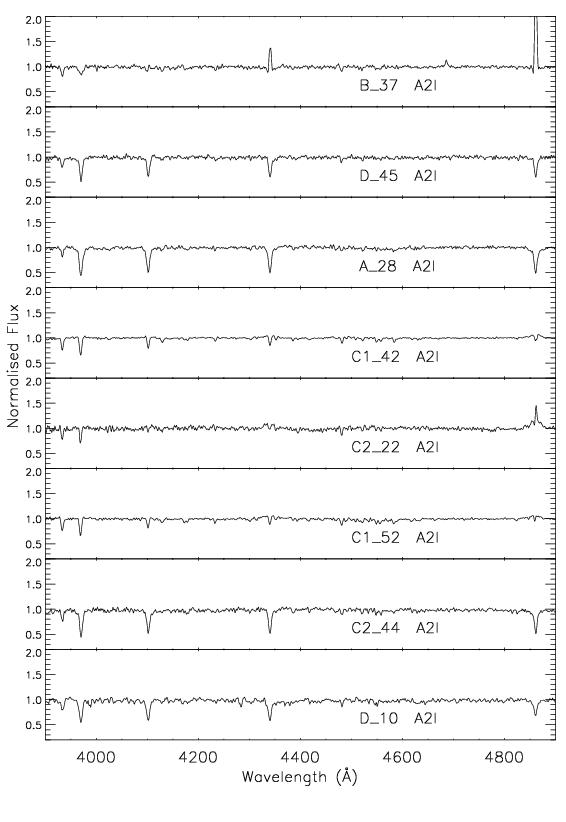}
\caption{A2 supergiants in NGC 55.}
\label{ape15}
\end{figure*}

\clearpage
\begin{figure*}
\includegraphics[width=18cm]{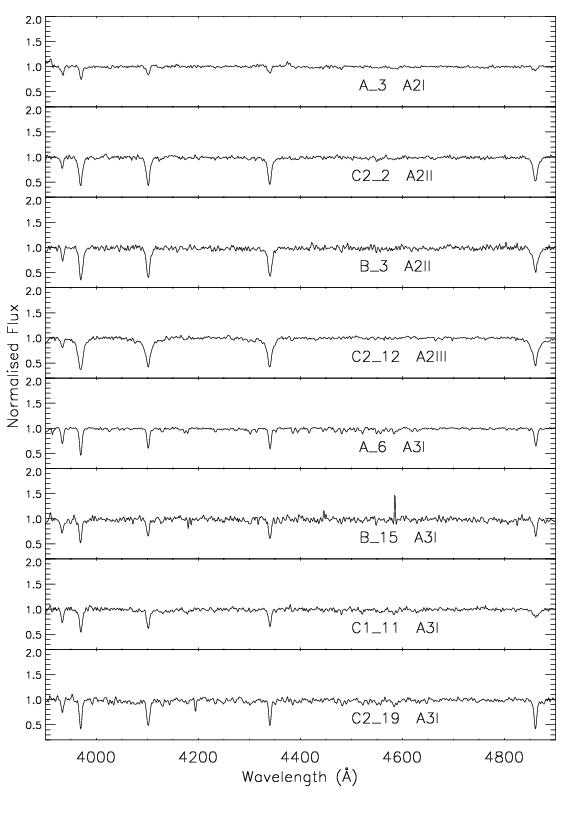}
\caption{A2 and A3 stars in NGC 55.}
\label{ape16}
\end{figure*}

\clearpage
\begin{figure*}
\includegraphics[width=18cm]{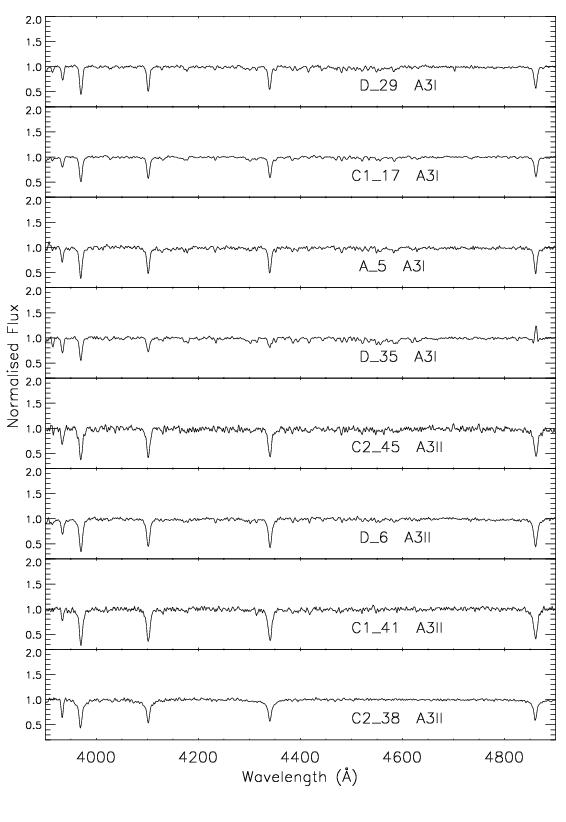}
\caption{A3 stars in NGC 55.}
\label{ape17}
\end{figure*}

\clearpage
\begin{figure*}
\includegraphics[width=18cm]{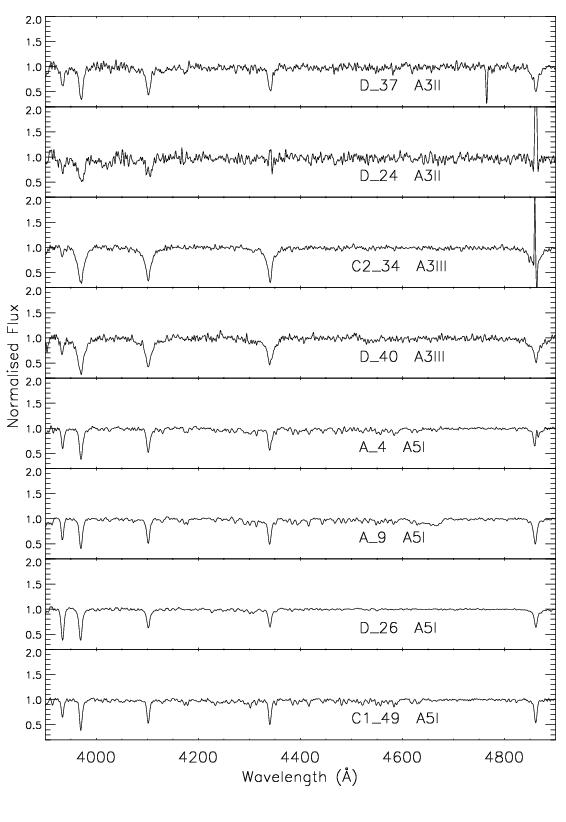}
\caption{Mid-A stars in NGC 55.}
\label{ape18}
\end{figure*}

\clearpage
\begin{figure*}
\includegraphics[width=18cm]{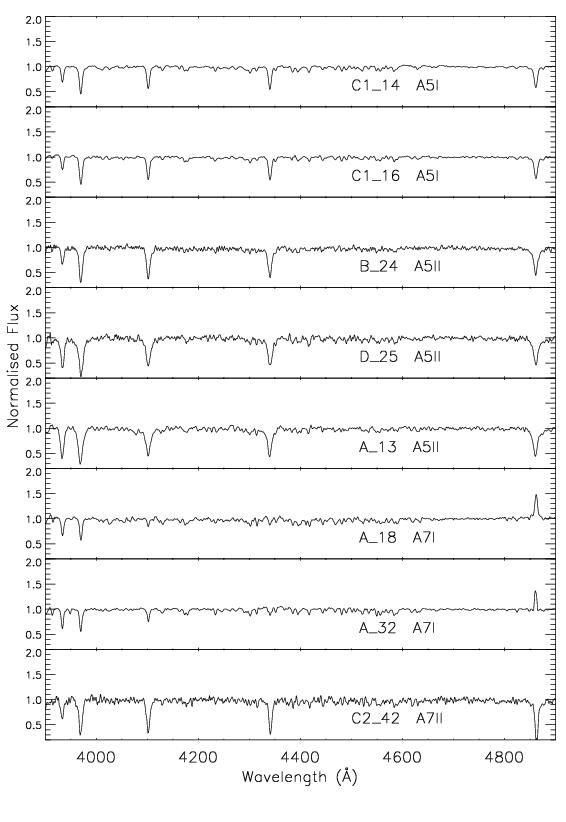}
\caption{Late-A stars in NGC 55.}
\label{ape19}
\end{figure*}

\clearpage
\begin{figure*}
\includegraphics[width=18cm]{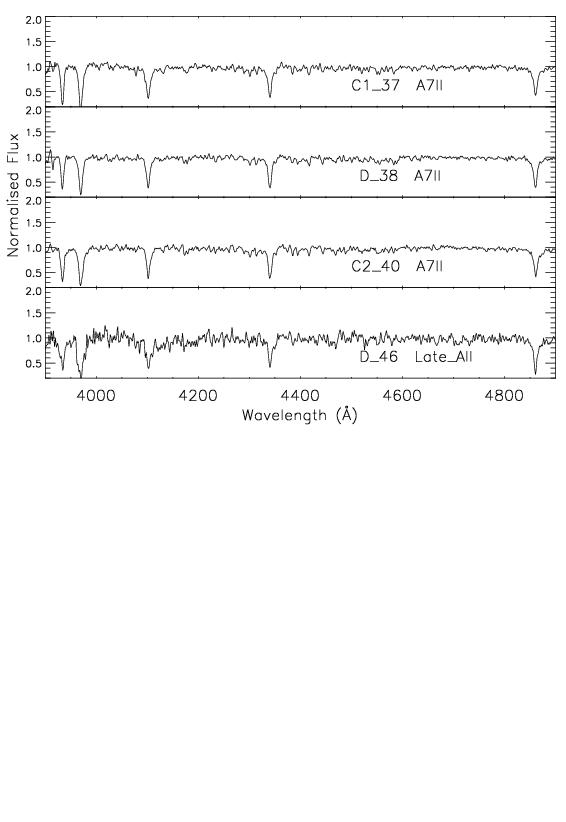}
\caption{Late-A stars in NGC 55, continued.}
\label{ape20}
\end{figure*}